\documentclass[tighten, twocolumn]{aastex701}

\usepackage{graphicx}	
\usepackage{amsmath}	
\usepackage{bm}
\usepackage{hyperref}
\usepackage{algorithm}
\usepackage[noend]{algpseudocode}
\usepackage[FIGTOPCAP]{subfigure}
\usepackage{tabularx}
\usepackage{booktabs}
\usepackage{threeparttable}
\usepackage{xspace}

\allowdisplaybreaks

\renewcommand{\vec}[1]{\boldsymbol{#1}}

\newcommand{\gaap}{\texttt{GAaP}}
\newcommand{\auto}{\texttt{AUTO}}
\newcommand{\noiselessfluxes}{\vec{F}}
\newcommand{\sigmaGAaP}{\vec{\sigma}_{\gaap}}
\newcommand{\sigmaAUTO}{\sigma_{r_{\auto}}}
\newcommand{\sigGAaPscalar}{\sigma_{r_{\gaap}}}
\newcommand{\noisyAUTOflux}{\hat{F}_{\auto}}
\newcommand{\noisyGAaPfluxes}{\hat{\vec{F}}_{\gaap}}
\newcommand{\detect}{\text{det}}
\newcommand{\lensfit}{\textit{lens}fit\xspace}

\newcommand{\shark}{\texttt{shark}\xspace}
\newcommand{\pc}{\texttt{pop-cosmos}\xspace}
\newcommand{\mlim}{\textit{Ch.\,1}\xspace}

\def\ln{\mathrm{ln}}

\graphicspath{{./}{figures/}}

\submitjournal{ApJ}

\shorttitle{Forward modeling KiDS with \pc}
\shortauthors{Leistedt et al.}

\begin{document}

\title{\pc: Forward modeling KiDS-1000 redshift distributions using realistic galaxy populations}

\correspondingauthor{Boris Leistedt}
\email{b.leistedt@imperial.ac.uk}

\author[0000-0002-3962-9274]{Boris Leistedt}
\affiliation{Astrophysics Group, Imperial College London, Blackett Laboratory, Prince Consort Road, London, SW7 2AZ, UK}
\email{b.leistedt@imperial.ac.uk}

\author[0000-0002-2519-584X]{Hiranya V.\ Peiris}
\affiliation{Institute of Astronomy and Kavli Institute for Cosmology, University of Cambridge, Madingley Road, Cambridge CB3 0HA, UK}
\affiliation{Cavendish Laboratory, Department of Physics, University of Cambridge, JJ Thomson Avenue, Cambridge, CB3 0HE, UK}
\affiliation{The Oskar Klein Centre, Department of Physics, Stockholm University, AlbaNova University Centre, SE 106 91 Stockholm, Sweden}
\email{hiranya.peiris@ast.cam.ac.uk}

\author[0000-0002-0352-9351]{Anik Halder}
\affiliation{Institute of Astronomy and Kavli Institute for Cosmology, University of Cambridge, Madingley Road, Cambridge CB3 0HA, UK}
\affiliation{Jesus College, Jesus Lane, Cambridge, CB5 8BL, UK}
\email{ah2425@cam.ac.uk}

\author[0009-0005-6323-0457]{Stephen Thorp}
\affiliation{Institute of Astronomy and Kavli Institute for Cosmology, University of Cambridge, Madingley Road, Cambridge CB3 0HA, UK}
\email{stephen.thorp@ast.cam.ac.uk}
\affiliation{The Oskar Klein Centre, Department of Physics, Stockholm University, AlbaNova University Centre, SE 106 91 Stockholm, Sweden}

\author[0000-0002-0041-3783]{Daniel J.\ Mortlock}
\affiliation{Astrophysics Group, Imperial College London, Blackett Laboratory, Prince Consort Road, London, SW7 2AZ, UK}
\affiliation{Department of Mathematics, Imperial College London, London SW7 2AZ, UK}
\email{d.mortlock@imperial.ac.uk}

\author[0000-0002-4371-0876]{Arthur Loureiro}
\affiliation{The Oskar Klein Centre, Department of Physics, Stockholm University, AlbaNova University Centre, SE 106 91 Stockholm, Sweden}
\affiliation{Astrophysics Group, Imperial College London, Blackett Laboratory, Prince Consort Road, London, SW7 2AZ, UK}
\email{arthur.loureiro@fysik.su.se}

\author[0000-0003-4618-3546]{Justin Alsing}
\affiliation{The Oskar Klein Centre, Department of Physics, Stockholm University, AlbaNova University Centre, SE 106 91 Stockholm, Sweden}
\email{justin.alsing@fysik.su.se}

\author[0009-0004-7935-2785]{Gurjeet Jagwani}
\affiliation{Institute of Astronomy and Kavli Institute for Cosmology, University of Cambridge, Madingley Road, Cambridge CB3 0HA, UK}
\affiliation{Research Computing Services, University of Cambridge, Roger Needham Building, 7 JJ Thomson Ave, Cambridge CB3 0RB, UK}
\email{gj329@cam.ac.uk}

\author[0000-0002-3287-1193]{Madalina N.\ Tudorache}
\affiliation{Institute of Astronomy and Kavli Institute for Cosmology, University of Cambridge, Madingley Road, Cambridge CB3 0HA, UK}
\email{mnt26@cam.ac.uk}

\author[0000-0003-1943-723X]{Sinan Deger}
\affiliation{Institute of Astronomy and Kavli Institute for Cosmology, University of Cambridge, Madingley Road, Cambridge CB3 0HA, UK}
\email{sd2062@cam.ac.uk}

\author[0000-0001-6755-1315]{Joel Leja}
\affiliation{Department of Astronomy \& Astrophysics, The Pennsylvania State University, University Park, PA 16802, USA}
\affiliation{Institute for Computational \& Data Sciences, The Pennsylvania State University, University Park, PA 16802, USA}
\affiliation{Institute for Gravitation \& the Cosmos, The Pennsylvania State University, University Park, PA 16802, USA}
\email{joel.leja@psu.edu}

\author[0009-0005-5575-1121]{Benedict Van den Bussche}
\affiliation{Institute of Astronomy and Kavli Institute for Cosmology, University of Cambridge, Madingley Road, Cambridge CB3 0HA, UK}
\email{bv322@cam.ac.uk}

\author[0000-0001-7363-7932]{Angus H. Wright}
\affiliation{Ruhr University Bochum, Faculty of Physics and Astronomy, Astronomical Institute (AIRUB), German Centre for Cosmological Lensing (GCCL), 44780 Bochum, Germany}
\email{awright@astro.rub.de}

\author[0000-0001-9952-7408]{Shun-Sheng Li}
\affiliation{Kavli Institute for Particle Astrophysics and Cosmology (KIPAC), 
SLAC National Accelerator Laboratory and Stanford University, Stanford, CA 94305, USA}
\affiliation{Leiden Observatory, Leiden University, PO Box 9513, 2300 RA Leiden, The Netherlands}
\email{liss@stanford.edu}

\author[0000-0002-4677-0516]{Konrad Kuijken}
\affiliation{Leiden Observatory, Leiden University, PO Box 9513, 2300 RA Leiden, The Netherlands}
\email{kuijken@strw.leidenuniv.nl}

\author[0000-0002-1194-6758]{Hendrik Hildebrandt}
\affiliation{Ruhr University Bochum, Faculty of Physics and Astronomy, Astronomical Institute (AIRUB), German Centre for Cosmological Lensing (GCCL), 44780 Bochum, Germany}
\email{hendrik.hildebrandt@ruhr-uni-bochum.de}

\begin{abstract}
\noindent
The accuracy of the cosmological constraints from Stage~IV galaxy surveys will be limited by how well the galaxy redshift distributions can be inferred.
We have addressed this challenging problem for the Kilo-Degree Survey (KiDS) cosmic shear sample
by developing a forward-modeling framework with two main ingredients: (1) the \texttt{pop-cosmos} generative model for the evolving galaxy population, calibrated on \textit{Spitzer} IRAC $\textit{Ch.\,1}<26$ galaxies from COSMOS2020; and (2) a data model for noise and selection, machine-learned from the SURFS-based KiDS-Legacy-Like Simulations (SKiLLS).
Applying KiDS tomographic binning to our synthetic photometric data, we infer redshift distributions in each of five bins directly from the population and data models, bypassing the need for spectroscopic reweighting.  Keeping the data model fixed, we compare results using two different galaxy population models: \texttt{pop-cosmos}; and \texttt{shark}, the semi-analytic galaxy formation model used in SKiLLS. In the first ($0.1<z<0.3$) and last ($0.9<z<1.2$) tomographic bins we find systematic differences in the mean redshifts of $\Delta z\sim0.05$-$0.1$,  comparable to the reported uncertainties from spectroscopic reweighting methods. This work paves the way for accurate redshift distribution calibration for Stage~IV surveys directly through forward modeling, thus providing an independent cross-check on spectroscopic-based calibrations which avoids their selection biases and incompleteness. We will use the \texttt{pop-cosmos} redshift distributions in an upcoming full KiDS cosmology reanalysis. 

\end{abstract}


\section{Introduction}

The extraction of cosmological information from galaxy surveys relies critically on our ability to model and understand the galaxy populations we observe. 
Modern wide-field surveys like the Kilo-Degree Survey (KiDS; \citealp{dejong2013}), Dark Energy Survey (DES; \citealp{des2005}), and Hyper Suprime-Cam Subaru Strategic Program (HSC SSP; \citealp{aihara2018}) provide unprecedented statistical power for weak gravitational lensing measurements. However, this power can only be fully realized by accurately characterizing the observed galaxy samples, in particular their redshift distributions, $n(z)$ \citep[see e.g.][]{mandelbaum2018, newman2022}.
Traditional approaches to this problem have relied on  empirical calibrations using spectroscopic samples.
While spectroscopic calibrations provide direct measurements, they suffer from incompleteness and their own selection biases. 
In this work, we reframe this as \emph{forward-modeling redshift calibration}: rather than directly re-weighting spectroscopic samples to estimate $n(z)$, we start from a generative model of the galaxy population that reproduces observed color–redshift relations, forward-model it through a realistic KiDS data model, and then apply exactly the same tomographic binning procedure as for the data. 

Image simulations are the ideal tool to forward model photometric data since they  offer complete control over the input population and can accurately capture all known observational effects.
The SURFS-based KiDS-Legacy-Like Simulations (SKiLLS; \citealp{Li_2023}) exemplify this approach: by injecting galaxies from the \shark semi-analytic model \citep{lagos2018} into synthetic images with the \textsc{GalSim} image simulation software \citep{rowe2015}, incorporating realistic point spread functions, backgrounds, and noise, SKiLLS enabled validation of the KiDS-1000 self-organizing map redshift calibration \citep{Hildebrandt2021, Wright2025} and the KiDS shear measurement pipelines\footnote{Image simulations have been also been used by other major weak lensing surveys to calibrate shear measurements and de-blending strategies (see e.g.\ \citealp{mandelbaum2018hsc, li2022hsc, maccrann2022}).} \citep{Li2023cosmicshear, yoon2025kids1000}. However, generating $\sim100$ deg$^2$ of mock images required a prohibitive amount of computing time. 
Image simulations have also been employed within forward modeling frameworks to constrain redshift distributions and luminosity functions \citep[e.g.][]{herbel2017, tortorelli2018, tortorelli2020, moser2024, fischbacher2025, fischbacher2024} using the Ultra Fast Image Generator (UFig; \citealp{berge2013, fischbacher2025_ufig}).

Recent advances in machine learning\footnote{Machine learning refers to computational methods that learn patterns from data without explicit programming. In our context, this includes generative models (synthesizing realistic galaxy properties) and supervised models (learning mappings between observables; e.g., noiseless to noisy photometry).} and generative modeling make it possible to construct data-driven population models that can be used to rapidly generate realistic galaxy catalogs while maintaining the flexibility for exploration of systematic uncertainties \citep[see][]{alsing2023}. This type of approach has been used to produce mock galaxy imaging \citep[e.g.][]{berge2013, herbel2017, moser2024, fischbacher2024, fischbacher2025, tortorelli2025}, provide GPU-based SPS libraries \citep[e.g.][]{hearin2023, alarcon2023, alarcon2025}, and generate forward models for spectroscopic data \citep[e.g.][]{hahn2023, hahn2024}. Our contribution to this field has been to develop the \pc framework \citep{alsing2024, thorp2024, thorp2025, deger2025}, which is the primary tool used in this paper. By training score-based diffusion models on deep photometric data from the Cosmic Evolution Survey (COSMOS; \citealp{scoville2007}), \pc encodes galaxies with realistic correlations between their physical properties and observed photometry. 
The model operates in the space of stellar population synthesis (SPS; see e.g.\ \citealp{conroy2013, iyer2026}) parameters, enabling direct inference of physical properties while maintaining consistency with observed color and magnitude distributions. 
Crucially, \pc achieves excellent photometric redshift performance (estimating galaxy distances from multi-band colors rather than spectra; \citealt{thorp2024, thorp2025}).

In this work, we develop a forward modeling framework for inferring the tomographic redshift distributions of galaxies in the KiDS-1000 weak lensing catalog \citep{kuijken2019, giblin2021}, based on the KiDS data and SKiLLS image simulations introduced in \autoref{sec:data}. 
This framework combines the \pc population model with a machine-learned data model, trained on  SKiLLS, that captures the transformation from noiseless galaxy photometry to realistic survey observations, including photometric noise, detection thresholds, and measurement uncertainties (described in detail in 
\autoref{sec:model}).
This allows us to rapidly generate mock KiDS observations for any input galaxy populations, including \pc and \shark.
We then apply our framework to estimate redshift distributions for the KiDS-1000 tomographic bins, demonstrating how they are affected by the choice of underlying galaxy population.
We discuss these results in 
\autoref{sec:discussion} and summarise our conclusions in \autoref{sec:conclusions}.
This work presents the first application of \pc to a galaxy survey beyond the COSMOS dataset on which it was originally calibrated, complementing A.\ Halder et al.\ (2026, hereafter H26), who apply \pc to individual KiDS-1000 galaxies for the inference of individual galaxy properties.

All magnitudes in this work are given in the AB system. The \pc\ model assumes a flat $\Lambda$CDM cosmology with $H_0=67.66$~km\,s$^{-1}$\,Mpc$^{-1}$ and $\Omega_{\textrm{m}}=0.3097$ \citep{planck18}, while SKiLLS assumes a flat $\Lambda$CDM cosmology with $H_0=67.51$~km\,s$^{-1}$\,Mpc$^{-1}$ and $\Omega_{\textrm{m}}=0.3121$ \citep{planck15}.

\section{Data}
\label{sec:data}

In this section we describe the KiDS DR4 data and the galaxy samples we consider (\autoref{sec:kids_data}). 
We also describe the SKiLLS image simulations (\autoref{sec:im_sims}).

\subsection{KiDS data}
\label{sec:kids_data}

The KiDS Data Release 4 (DR4; \citealp{kuijken2019}) covers 1022 deg$^2$ of optical imaging in four bands ($u g r i$) obtained with the OmegaCAM wide-field imager \citep{kuijken2002, kuijken2011} on the VLT Survey Telescope (VST; \citealp{arnaboldi1998, capaccioli2011}). 
The survey footprint consists of two main regions: KiDS-North ($\sim$450 deg$^2$) overlapping with the Galaxy and Mass Assembly (GAMA) survey \citep{driver2009}; and KiDS-South ($\sim$550 deg$^2$) overlapping with the 2 degree Field Galaxy Redshift Survey (2dFGRS; \citealp{colless2001}). 
The median seeing ranges from 0.70$''$ in $r$-band to 1.00$''$ in $u$-band. The $r$-band limiting magnitude is $\sim$25.0 ($5\sigma$ in a 2$''$ aperture).
KiDS DR4 provides photometry for $\sim 10^8$ sources using two complementary approaches: Gaussian Aperture and PSF (\gaap; \citealp{kuijken2008}) photometry for accurate colors; and Source Extractor \citep{bertin1996} \auto\ magnitudes for fluxes. 
Detection and star-galaxy separation are performed in the $r$-band, with sources required to have signal-to-noise ratio greater than 5 to be considered as detected. The nine-band ($ugriZYJHK_{\rm s}$) photometric catalog is created by combining KiDS optical data with near-infrared data from the VISTA Kilo-degree INfrared Galaxy survey (VIKING; \citealp{edge2013}), enabling improved photometric redshift estimation \citep{Hildebrandt2021}.

Weak-lensing studies use the KiDS-1000 source sample \citep{giblin2021}, which consists of $\sim 2.1\times10^6$ galaxies from DR4 with robust shape
measurements obtained with \lensfit \citep{miller2013}. These galaxies are divided into five tomographic bins using Bayesian Photo-$Z$ (BPZ; \citealp{benitez2000}) photometric redshifts \citep{wright2019}, with bin edges at redshifts of 0.1, 0.3, 0.5, 0.7, 0.9 and  1.2. 
Additional quality cuts remove spurious detections near bright stars, in masked regions, and at survey boundaries, resulting in an effective area of 777.4 deg$^2$ for cosmological analyses.

Unfortunately, some of the information required to construct the KiDS-1000 selection step-by-step from the DR4 catalog is not publicly available. Therefore, to support our analysis, we cross-matched the KiDS DR4 public catalogs with the KiDS-1000 science catalog to obtain the KiDS-1000 flag, a binary indicator of which galaxies pass the selection criteria for weak lensing analysis, along with their associated \lensfit weight \citep{giblin2021}\footnote{Additionally, we cross-matched with internal (non-public) catalogs, kindly provided by the KiDS team, which contain photometric uncertainties for galaxies with negative measured fluxes. These were not preserved in the ESO public DR4, but were essential for reproducing both the selection function and the complete noise properties of the KiDS photometry. 
In particular, to handle missing data in the DR4 catalogs, which appear flagged by $99$s (flux signal-to-noise ratio lower than one) or $-99$s (no data), we implemented a resampling strategy that draws missing flux values from the flux uncertainty distributions. This ensures physically plausible imputation while maintaining the statistical properties of the data. This limitation of the ESO database was fixed for DR5.}.

The KiDS DR4 catalog applies corrections for Galactic extinction using the \citet{schlegel1998} $E(B-V)$ maps with \citet{schlafly2011} extinction coefficients, and photometric zero-point calibrations tied to Gaia DR2 using stellar locus regression \citep{kuijken2019}. 
These corrections are already incorporated into the released \gaap\ magnitudes, which are provided as dereddened, calibrated apparent magnitudes. 
However, the catalog employs a two-magnitude system that requires aperture correction to recover total galaxy fluxes. The \auto\ magnitude in the $r$-band is Kron-like \citep{kron1980, bertin1996} and approximates the total light of galaxies, while the \gaap\ magnitudes across all nine bands ($ugriZYJHK_{\rm s}$) are aperture-matched and PSF-corrected across bands. 
The latter provide accurate colors, but underestimate total fluxes for extended sources (since the aperture size is constrained to avoid contamination from neighboring objects; \citealp{kuijken2008, kuijken2019}). 
To obtain total flux estimates in bands other than $r$, we apply the aperture correction by combining the $r$-band \auto\ magnitude with \gaap\ colors \citep{wright2019}.  In band $b \in \{u, g, r, i, Z, Y, J, H, K_{\rm s}\}$ this gives
\begin{equation}
    \hat{F}_{b_\mathrm{\gaap}}^\mathrm{corrected} =
    \hat{F}_{b_\mathrm{\gaap}}
       \frac{\hat{F}_{r_\mathrm{\auto}}} {\hat{F}_{r_\mathrm{\gaap}}} .
\end{equation}
The resultant fractional uncertainty is given by the quadrature sum  
\begin{equation}
\left(\frac{\sigma_{b_\mathrm{\gaap}}^\mathrm{corrected}}{\hat{F}_{b_\mathrm{\gaap}}^\mathrm{corrected}}
    \right)^2 = \left(\frac{\sigma_{b_\mathrm{\gaap}}}{\hat{F}_{b_\mathrm{\gaap}}}\right)^2 + \left(\frac{\sigmaAUTO}{\hat{F}_{r_\mathrm{\auto}}}\right)^2 + \left(\frac{\sigGAaPscalar}{\hat{F}_{r_\mathrm{\gaap}}} \right)^2.
\end{equation}
This correction procedure assumes no color gradients within galaxies, an assumption which has been tested by comparing \gaap\ fluxes measured with different aperture sizes \citep{kuijken2019}.
In our approach, we jointly model the $r$-band \auto\ magnitude and the $ugriZYJHK_{\rm s}$ aperture-corrected \gaap\ magnitudes.
This choice is supported by the robust photometric redshifts obtained by H26 by analysing the same photometry through SED fitting under the \pc prior\footnote{For bright galaxies the combination of \gaap\ colors and Kron‑like \auto\ magnitudes can lead to a modest underestimation of the total flux for very extended sources \citep{graham2005}, such as luminous red galaxies (LRGs). This effect has been investigated in detail in the companion analysis of H26, who show that it can induce small but measurable biases in photometric redshifts for DESI LRGs when using KiDS photometry alone (see section 4.1 of H26). In the present work, however, we focus on the KiDS‑1000 weak‑lensing source sample, which is dominated by fainter and less extended galaxies near the survey flux limit and strongly down‑weighted by the \lensfit scheme at high luminosities. Consequently, the specific \gaap$-$\auto~flux bias that affects bright LRGs is expected to have a negligible impact on the tomographic source redshift distributions.}.

\subsection{Image simulations}
\label{sec:im_sims}

The SKiLLS simulations \citep{Li_2023} provide realistic mock observations designed to match the properties of the KiDS data (building on the earlier KiDS simulations by \citealp{fenechconti2017} and \citealp{kannawadi2019}). 
These simulations combine cosmologically representative galaxy populations with observationally calibrated morphologies to create synthetic images that capture the complexity of the survey selection function.
The galaxy population underlying SKiLLS comes from the SURFS--\shark framework, which couples $N$-body simulations from the Synthetic UniveRses For Surveys (SURFS; \citealp{elahi2018}) project with the \shark semi-analytic galaxy formation model \citep{lagos2018}.
\shark predicts galaxy properties including stellar masses, star formation histories, metallicities, and active galactic nucleus (AGN) activity, based on physical prescriptions for gas cooling, star formation, feedback, and mergers. 

Multi-band photometry is generated using the ProSpect SPS code \citep{robotham2020prospect}, with the \cite{charlot2000} dust attenuation model  and \cite{dale2014} dust emission templates.
A key feature of SKiLLS is the incorporation of realistic galaxy morphologies learned from Hubble Space Telescope (HST) observations in the COSMOS field \citep{griffith2012}. 
This is based on vine copulas \citep[see e.g.][]{joe2014,czado2019} to capture the  correlations between morphological parameters (size, ellipticity, S\'{e}rsic index) and other galaxy properties (magnitude, color, redshift), extending to magnitudes and redshifts beyond the HST sample.
The image simulations inject these galaxies into realistic KiDS-like images, including PSFs, sky backgrounds, and noise properties matched to the survey data \citep{Li_2023}. 
Galaxies are distributed with realistic clustering following the dark matter halos from the SURFS $N$-body simulations \citep{elahi2018}. 
An empirical correction based on stellar mass-to-light ratios ensures the final galaxy number counts match observations in the COSMOS2015 dataset \citep{davidzon2017cosmos2015}. 
The simulated images are processed through the same pipeline used for real KiDS data, resulting in realistic catalogs with identical data products, including \gaap\ and \auto\ photometry, shape measurements, and photometric redshifts.

The SKiLLS noise model assumes Gaussian distributions for pixel noise. For optical bands, the noise properties are calibrated from Astro-WISE weight maps \citep{McFarland2013} with an empirical boost factor of $\sim1.145$ to account for the re-gridding process \citep{Li_2023}. 
For the VIKING infrared bands the noise modeling is simplified: rather than simulating the complex multi-exposure paw-print structure, SKiLLS uses single images per tile with noise levels derived by stacking flat-field images from overlapping paw-prints and taking median values. This simplified approach maintains realistic photometric properties while reducing computational overhead. The noise model has been validated by comparing the simulated \gaap\ limiting magnitudes to data, showing agreement within $0.1$ mag for optical bands and somewhat larger scatter for NIR bands \citep{Li_2023}.

For this work, we use both the output measurement catalogs containing $2\times10^6$ detected galaxies, and the input truth catalogs containing the noiseless properties of all injected galaxies. 
The availability of matched input--output pairs enables us to train our data model on the complex transformation between noiseless fluxes and noisy data, including detection probabilities, photometric scatter, and measurement biases.
The simulations cover 100 deg$^2$ with depth variations representative of the full KiDS footprint, providing a comprehensive training set for our data model.
To ensure a consistent treatment of noisy measurements, we apply to the simulated low signal-to-noise fluxes exactly the same resampling algorithm that we used for the real DR4 data, as described in the previous section.

\section{Modeling}
\label{sec:model}

Our forward model for KiDS consists of five ingredients, which are applied in succession when constructing mock redshift distributions for the tomographic bins:
{a population model} (\autoref{sec:pop-model})
; {a data model} (\autoref{sec:data-model})
; {selection and weighting} (Sections \ref{sec:k1k-selection}  and \ref{sec:weight-prediction});
and {a photometric redshift estimator} (\autoref{sec:tomo-selection}).
Taken together, these ingredients define our forward-modeling redshift calibration: they allow us to start from a generative model of the galaxy population, create KiDS-like mock observations that pass the same tomographic binning as the data, and then read off the underlying true redshift histograms of the simulated galaxies as calibrated estimates of the KiDS-1000 tomographic $n(z)$.

\subsection{Galaxy population model}
\label{sec:pop-model}   
Our analysis is based on the \pc generative model for the redshift-evolving galaxy population \citep{alsing2024, thorp2025}. This is an SPS-based model, trained on deep ultraviolet (UV) to mid-infrared (MIR) photometry in 26 broad and narrow bands from the COSMOS2020 catalog \citep{weaver2022}. The \pc training set consists of $\sim420,000$ MIR-selected galaxies with Spitzer IRAC $\mlim<26$. 
The SPS parametrization in \pc is based on a variant of the Prospector-$\alpha$ model \citep{leja2017, leja2018, leja2019, leja2019_sfh}, providing a 16-parameter representation of each galaxy. The model uses: a non-parametric star formation history (SFH; \citealp{leja2019_sfh}), represented by 6 parameters; a two-component dust attenuation model \citep{charlot2000}, including a birth-cloud (affecting stars younger than $10$~Myr) and diffuse (affecting all stars) component; a variable attenuation law slope and associated UV bump strength \citep{calzetti2000, noll2009, kriek2013}; and an IR active galactic nucleus (AGN) contribution, associated with a clumpy dust torus \citep{nenkova2008i, nenkova2008ii, leja2018}. The ionization parameter and gas-phase metallicity can vary independently from stellar metallicity, and nebular emission (lines and continuum) is based on the \texttt{CLOUDY} \citep{ferland2013} photoionization code (via model grids from \citealp{byler2017}).  The 16 SPS parameters used in \pc are listed in \autoref{tab:sps_parameters}.

\begin{table}
    \centering
    \caption{The SPS and other parameters used in the \pc galaxy population model.}
    \label{tab:sps_parameters}
    \begin{tabular}{l l}
        \toprule \toprule
        symbol / unit & definition\\
        \midrule
        $\log_{10}(M^\text{form}_\star/\mathrm{M}_\odot)$ & logarithm of stellar mass formed\\
        $\log_{10}(Z_\star/\mathrm{Z}_\odot)$ & logarithm of stellar metallicity \\
        $\Delta\log_{10}(\text{SFR})_{\{2:7\}}$ & SFR ratios between SFH bins \\
        $\tau_2/\text{mag}$ & diffuse dust optical depth \\
        $n$  & index for diffuse dust attenuation law \\
        $\tau_1/\tau_2$ & birth cloud dust optical depth \\
        $\ln(f_\text{AGN})$ & logarithm of AGN luminosity fraction \\
        $\ln(\tau_\text{AGN})$ & logarithm of AGN torus optical depth \\
        $\log_{10}(Z_\text{gas}/\mathrm{Z}_\odot)$ & logarithm of gas-phase metallicity \\
        $\log_{10}(U_\text{gas})$ & logarithm of gas ionization\\
        $z$ & redshift \\
        \bottomrule
    \end{tabular}
\end{table}

The SPS calculations in \pc are accelerated using the \texttt{Speculator} emulator \citep{alsing2020}, a neural network model that is trained to reproduce the outputs of the Flexible Stellar Population Synthesis \citep[FSPS;][]{conroy2009, conroy2010, conroygunn2010} and Prospector \citep{johnson2021} libraries to better than $\sim3\%$ precision in flux across 99.9\% of allowable parameter space, and to better than $\sim1\%$ precision across 99\% of parameter space. For this work, we have trained emulators to predict FSPS/Prospector photometry for the KiDS passbands: the VST/OmegaCAM $ugri$ bands and the VISTA/VIRCAM $ZYJHK_{\rm s}$ bands.

To describe the galaxy population, \pc defines a 16-dimensional probability distribution over the SPS parameters, represented using a score-based diffusion model \citep{song21}. Mock galaxies can be generated by taking draws of SPS parameters from the diffusion model, passing these through the \texttt{Speculator} emulator described above to generate noiseless photometry, incorporating recalibration terms for nebular emission and photometric zero points (using the formalism from \citealp{leistedt2023hierarchical}), and then applying a survey-specific data-model and selection effects to produce realistic noisy photometry for any survey with compatible selection effects (i.e., not deeper than the $\mlim<26$ COSMOS2020 catalog used to train the model). The sections that follow will describe the construction of these models for KiDS, for rapid synthetic galaxy catalog generation ($\sim$1 million galaxies$/$GPU-hour). 

As well as being used in a predictive capacity, the learned population distribution can be deployed as a prior over SPS parameters, to be used in SED fitting of individual galaxies' photometry. This provides additional leverage over less-informative broader priors \citep[e.g.,][]{leja2017}, and offers a data-driven alternative to priors based on empirical scaling relations \citep[e.g.,][]{wang2023}. \citet{thorp2024, thorp2025} demonstrated that when used as a data-driven prior in this way, \pc yields more accurate individual photometric redshifts than conventional SED fitting algorithms such as LePhare \citep{arnouts1999, ilbert2006, ilbert2009} and EAZY \citep{brammer2008}. 
In H26 a similar demonstration will be provided for redshifts derived from $ugriZYJHK_{\rm s}$ photometry of the KiDS-1000 galaxies with \pc, demonstrating the ability to go from the 26-band setting of \citet{thorp2024} to 9 bands.
In this work, we will be demonstrating \pc as a tool for estimating population-level redshift distributions.

The population model presented in \citet{thorp2025} is ideal for the task of estimating KiDS redshift distributions. The $\mlim<26$ COSMOS2020 training sample has a high degree of completeness at the KiDS limiting magnitude (based on COSMOS2020 data, we estimate that $\sim96\%$ of galaxies with $r<25$ would be included in an $\mlim<26$ selection). With appropriate selection applied, we therefore expect synthetic galaxies drawn from \pc\ to be representative of the galaxy population probed by KiDS. The first step to generate our KiDS-like mock catalogs is to repeatedly draw SPS parameters and redshifts from the \pc\ model, which are then passed through our KiDS SPS emulator to obtain noiseless model fluxes,  $\noiselessfluxes = (F_u, F_g, \ldots, F_{K_{\rm s}})$, in in the $ugriZYJHK_{\rm s}$ bands. 

\subsection{Data model}
\label{sec:data-model}

Taking our large sample of mock galaxies with noiseless model fluxes as an input, we construct a data model for the measurement noise and selection criteria which combine to generate the KiDS catalogs (Section~\ref{sec:kids_data}). Conceptually, the natural sequence for a given galaxy would be to i) start with the depth or noise level in each band, ii) add photometric noise to the noiseless model fluxes to obtain measured fluxes, and then iii) apply the selection criteria. We cannot do this, however, as the pre-existing SKiLLS image simulations \citep{Li_2023} from which we learn these distributions include photometric measurements only for detected sources. For this reason, we factorize the data model to use distributions conditioned on detection from the outset, without any loss of generality\footnote{In \cite{alsing2024} and \cite{thorp2025} we were able to avoid the complication of having to define a data model conditioned on detection as we used selection cuts of $r<25$ and $\mlim<26$, respectively.}.

 The quantities used in our KiDS data model, defined in Table~\ref{tab:symbols}, include both fluxes (most natural for the statistical formalism) and magnitudes (used for learning the relevant distributions). The $r$ band is singled out as the KiDS detection band; quantities in the other bands only need to be considered for detected sources. We characterise the depth of the images using the reported limiting magnitudes, which in KiDS are equivalent to the 1-sigma uncertainty\footnote{Throughout this work, limiting magnitudes refer to $1\sigma$ \gaap\ flux uncertainties rather than the $5\sigma$ detection thresholds more commonly adopted in optical surveys. This convention is natural for \gaap\ photometry, where aperture-matched flux uncertainties directly characterize measurement precision. We note an important distinction: while survey depth at a given position can be characterized by a limiting magnitude (determined by sky background, exposure time, and seeing), individual source flux uncertainties only approach this depth-based limit for faint sources in the background-dominated regime \citep{howell2006}. Brighter sources have additional uncertainty contributions from source photon statistics. This is particularly relevant for our data model, where we condition on both noiseless source fluxes and position-dependent depth, assuming background-limited uncertainties.} on the 9 \gaap\ fluxes, $\sigmaGAaP$.
Additionally, the uncertainty on the \auto\ flux, $\sigmaAUTO$, must also be modeled, because it differs from $\sigGAaPscalar$ given the different type of flux  (Section~\ref{sec:kids_data}). For a galaxy with noiseless model fluxes $\noiselessfluxes$ observed with \gaap\ uncertainties $\sigmaGAaP$ our data model is then defined by 
$p(\noisyAUTOflux, \noisyGAaPfluxes, \sigmaAUTO, \detect |  \sigmaGAaP, \noiselessfluxes)$, the joint distribution over detection, det, measured $r$-band \auto\ flux, $\noisyAUTOflux$ and the measured \gaap\ fluxes. We then factor this full distribution as 
\begin{eqnarray}\label{eq:datamodel}
p\left(\noisyAUTOflux, \noisyGAaPfluxes, \sigmaAUTO, \detect \Bigl|  \sigmaGAaP, \noiselessfluxes   \right)
=\nonumber\\
\underbrace{p\left(\detect \Bigl|  \sigmaGAaP, \noiselessfluxes  \right)}_{\mathrm{detection}} 
\times
\underbrace{p\left(\sigmaAUTO \Bigl| \sigmaGAaP, \noiselessfluxes, \detect \right)}_{\mathrm{uncertainty}} \nonumber\\
\times
\underbrace{p\left(\noisyAUTOflux, \noisyGAaPfluxes \Bigl| \sigmaAUTO, \sigmaGAaP, \noiselessfluxes, \detect    \right)}_{\mathrm{noise}},
\end{eqnarray}
where the first term is the detection model (Section~\ref{sec:detection-model}) and the second and third terms, which correspond to the uncertainty model (Section~\ref{sec:uncertainty-model}) and the noise model (Section~\ref{sec:noise-model}), are both conditioned on detection and so can be learned from the SKiLLS simulations.

\begin{table}
    \centering
    \caption{Symbol definitions for the data model. All magnitudes and fluxes are extinction- and aperture-corrected as described in the text.
    The first block lists quantities for a single band (shown for the $r$-band); the same notation applies to other bands by substituting the band name.
    The second block lists multi-band model quantities and derived indicators.
    }
    \label{tab:symbols}
    \begin{tabular}{p{1cm} p{6.5cm}}
        \toprule \toprule
symbol & definition \\
        \midrule
$r$ & $r$-band model (noiseless, FSPS) magnitude \\ 
$\sigGAaPscalar$ & \gaap\ $r$-band 1-sigma flux uncertainty \\
$r_\mathrm{\gaap, 1\sigma}$ &  $\sigGAaPscalar$ as an AB magnitude (magnitude limit) \\
$\hat{r}_\mathrm{\gaap}$ & \gaap\ $r$-band measured magnitude  \\
$\hat{F}_{r_\mathrm{\gaap}}$ & \gaap\ $r$-band measured flux  \\ \hline
$r_\mathrm{\auto}$ & \auto\ $r$-band measured magnitude\\
$\sigmaAUTO$ & \auto\ $r$-band 1-sigma flux uncertainty  \\
$r_{\mathrm{\auto},{1\sigma}}$ & $\sigmaAUTO$ as an AB magnitude  \\
$\hat{F}_{r_\mathrm{\auto}}$ & \auto\ $r$-band measured flux\\

$\detect$ & Detection boolean indicator \\
$\vec{F}$ & $ugriZYJHK_{\rm s}$ FSPS model fluxes \\
$\noisyAUTOflux$ & $r$-band \auto\ measured flux\\
$\noisyGAaPfluxes$ & $ugriZYJHK_{\rm s}$ \gaap\ noisy fluxes \\
$\sigmaGAaP$ & $ugriZYJHK_{\rm s}$ \gaap\ flux uncertainties (equivalent to the magnitude limits)\\
        \bottomrule
    \end{tabular}
\end{table}

\subsubsection{Detection model}
\label{sec:detection-model}

The KiDS catalogues are selected based on $r$-band  images, so the detection probability, written in generality as $p(\detect |  \sigmaGAaP, \noiselessfluxes)$ in Equation~\ref{eq:datamodel}, can be written more simply as $p(\detect | \sigmaGAaP, r)$, where the switch from flux to magnitude makes fitting easier. The difficulty here is that the actual detection method is not a simple magnitude/flux cut, but depends on factors inherent to the images, such as the number of connected pixels with values above a detection threshold.

We use a random forest classifier \citep{breiman2001random, scikit-learn} to represent $p(\detect | \sigmaGAaP, r)$, training it on 800,000 objects randomly sampled from SKiLLS truth input catalog.  For detected objects the inputs are the reported $r$-band magnitude and magnitude limit, while for undetected objects we assign magnitude limits drawn from the rest of the catalog.

We also need to ensure that the distributions of sampled $r$ magnitudes and the magnitude limits match the equivalent distributions in the KiDS data.
We implement this with a rejection sampling strategy.
We model the target distribution (obtained using the SKiLLS sample) with a second random forest, also trained on 800,000 sources.

\begin{figure}
    \centering
    \includegraphics[width=1\linewidth]{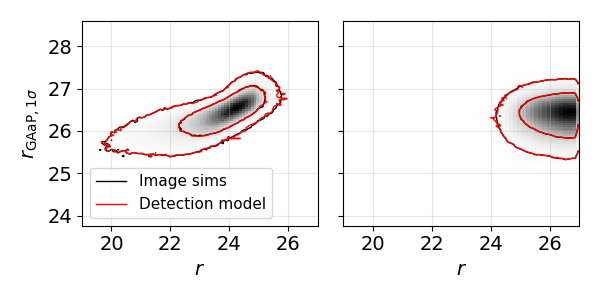}
    \caption{Validation of the $r$-band detection model: probability density of the \gaap\ magnitude limit given the noiseless model magnitude, for detected (left) and undetected (right) objects, for the SKiLLS simulations (black) and our model (red, described in described in \autoref{sec:detection-model}). The contours enclose 68\% and 95\% of the total probability.}
    \label{fig:detection-model}
\end{figure}

\autoref{fig:detection-model} shows a numerical validation of this two-step detection model, confirming that our model reproduces these distributions well.

\subsubsection{Uncertainty model}
\label{sec:uncertainty-model}

In KiDS the reported \gaap\ magnitude limits, corresponding to a \gaap\ flux of $1\sigma$, are a reasonable proxy for the depth and so do not need to be included in the data model.  However, this is not the case for the Kron-like \auto\ magnitude, and its error $\sigmaAUTO$, which sensitive to galaxy morphology (light profile, size, Sérsic index) and local conditions (PSF, neighboring sources). Since \pc does not provide a generative model for images, we need to use the SKiLLS image simulations learn the distribution $p\left(\sigmaAUTO | \sigmaGAaP, \noiselessfluxes, \detect \right)$ introduced in Equation~\ref{eq:datamodel}.

\begin{figure}
    \centering
    \includegraphics[width=1\linewidth]{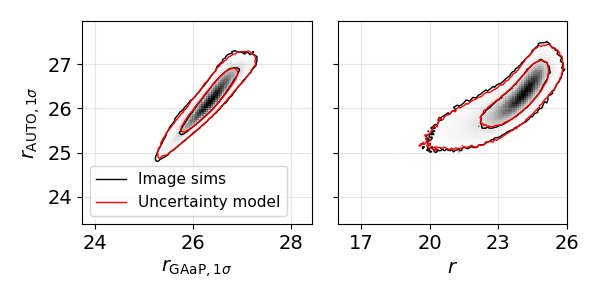}
    \caption{Validation of the $r$-band uncertainty model: probability density of the \auto\ flux uncertainty (vertical axes, converted to AB magnitude) as a function of the magnitude corresponding to a flux of $\sigGAaPscalar$ (left) and the noiseless model magnitude (right), for the SKiLLS simulations (black) and our model (red, described in \autoref{sec:uncertainty-model}). The contours enclose 68\% and 95\% of the total probability.}
    \label{fig:uncertainty-model}
\end{figure}

To learn $p\left(\sigmaAUTO | \sigmaGAaP, \noiselessfluxes, \detect \right)$ we use a conditional ODE flow \citep{lipman2023flow} trained with flow-matching\footnote{Our implementation, \texttt{flowfusion}, is publicly available at \url{https://github.com/Cosmo-Pop/flowfusion}.} on 800,000 training samples from SKiLLS, as detailed in \autoref{sec:uncertainty_architecture}. As illustrated in  \autoref{fig:uncertainty-model} this model is able to accurately reproduce the details of this distribution.

\subsubsection{Noise model}
\label{sec:noise-model}

Measured fluxes differ from true fluxes for several distinct reasons; these are jointly referred to as (photometric) noise, defined here by the distribution $p(\noisyAUTOflux, \noisyGAaPfluxes | \sigmaAUTO, \sigmaGAaP, \noiselessfluxes, \detect)$, as introduced in Equation~\ref{eq:datamodel}.
It is convenient to work in terms of residuals, i.e.\ the difference between the observed and model fluxes, divided by the uncertainty.
In principle, such residuals should be exactly uncorrelated if background subtraction uncertainties dominate and the time interval between the measurements being made is sufficiently long, as is the case here.
For ideal flux measurement and depth characterization pipelines, one would also expect these residuals to be nearly zero-mean, unit-variance Gaussian distributed, and independent of depth or input magnitudes. In practice, heavy tails and magnitude-dependent trends (and non-Gaussianity more generally) in these distributions are not only unavoidable, but are prevalent for the faint galaxies entering the KiDS-1000 sample. For example, galaxies near the detection limit show asymmetric errors with negative flux measurements, and faint extended sources exhibit systematically underestimated fluxes. These non-Gaussian features arise from source blending, background subtraction uncertainties, and the discrete pixel nature of images. Neglecting them would create discrepancies between the data and our simulated samples, and in turn, in the estimated redshift distributions.

For these reasons, we model the 10-dimensional flux noise residuals with another conditional ODE flow. 
The training data consists of 100 random tiles from SKiLLS \citep{Li_2023}, each residual being calculated as the difference between the reported flux measurement (\auto\ or \gaap) and the noiseless flux model (originally from the ProSpect SPS model; \citealp{robotham2020prospect}) divided by the reported flux uncertainty.
We also condition on the 9-band noiseless magnitudes.
Details of the training and model architecture are provided in \autoref{sec:noise_architecture}.

\begin{figure}
    \centering
    \includegraphics[width=1\linewidth]{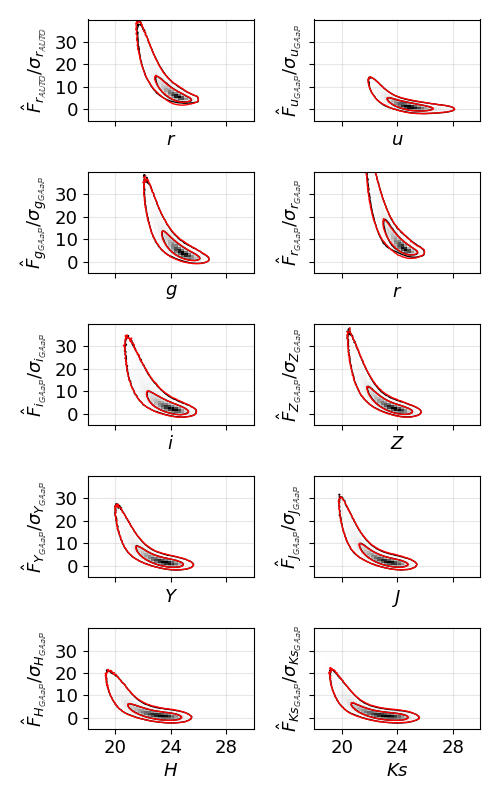}
    \caption{Validation of the noise model: probability density of the flux signal-to-noise ratio as a function of the noiseless model magnitude in each band, for the SKiLLS simulations (black) and our model (red, described in \autoref{sec:noise-model}). Contours enclose 68\%  and 95\%  of the joint probability density.}
    \label{fig:noise-model}
\end{figure}

\autoref{fig:noise-model} shows the resulting model, compared with the input data, demonstrating satisfactory agreement.
We found that alternative models such as normalizing flows or correlated Student's $t$ distributions did not perform well, especially struggling to perform simultaneously well at the bright and the faint ends of the flux noise residuals. 

\subsection{Tomographic binning}
\label{sec:tomo-selection}

For cosmological analyses, the
detected galaxies are commonly sorted into tomographic redshift bins. 
This separation can be achieved with any method, but the choice affects the constraining power and residual systematic biases (e.g., which redshift bin is more likely to capture the stars misclassified as galaxies, and how much this depends on the multi-band photometric noise).
In KiDS, photometric redshift estimates are based on BPZ \citep{benitez2000}. 
A posterior distribution is obtained by first evaluating a likelihood taking in the flux and flux uncertainties available, as a function of three parameters: redshift, absolute magnitude, and index of the galaxy SED template considered.
Second, a prior (a simple parametric function of redshift) is incorporated. BPZ then marginalizes over the absolute magnitude and the choice of SED template, in order to deliver a posterior probability distribution on redshift alone. 
The final estimated photometric redshift value, $Z_B$, is the maximum a posteriori (MAP) value, and in KiDS it is used to sort galaxies into redshift bins.
To keep as close to the KiDS team's analysis configuration as possible, we employ the BPZ configuration taken from the DR5 analysis \citep{wright2024} because we were unable to exactly reproduce the publicly released DR4 BPZ redshifts using the settings recommended in the KiDS analysis \citep{kuijken2019}. This could be due to untracked changes in configuration or input files.
The median and 95th-percentile of absolute differences in $Z_B$ between the two configurations are $0.01$ and $0.69$, respectively, indicating that differences are only significant for a small fraction of the catalog.
However, quantifying the net impact on $n(z)$ would require running our full forward-modeling pipeline with both setups, which was not possible. 
Because we use the DR5 BPZ configuration whereas the published KiDS-1000 $n(z)$ were derived with the DR4 setup, our tomographic samples are not strictly identical across the source sample, and differences between the two sets of $n(z)$ should be interpreted with this in mind. 

\subsection{KiDS-1000 selection}
\label{sec:k1k-selection}

\begin{figure*}
    \centering
    \includegraphics[width=0.99\linewidth]{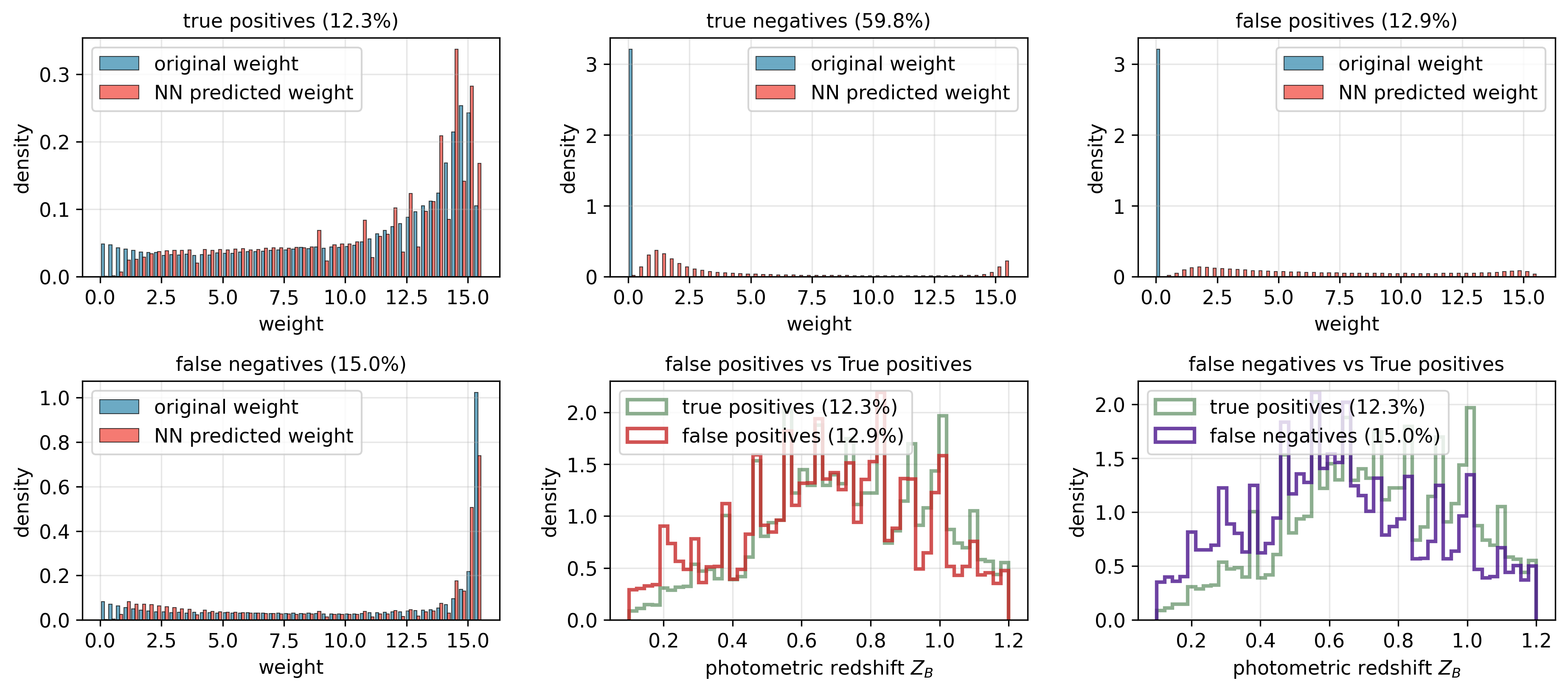}
    \caption{Performance of the classifier predicting the KiDS-1000 class, as a function of the \lensfit weight (true and predicted with the regressor presented in \autoref{sec:weight-prediction}) and the photometric redshift. True positives show higher \lensfit weights (left panel), confirming the classifier identifies high-quality sources. False positives cluster at low weights and high BPZ photometric redshifts $Z_B$, indicating contamination from low-quality measurements at the survey edges. }
    \label{fig:classifier_performance}
\end{figure*}

The KiDS-1000 cosmic shear sample \citep{giblin2021} is a curated subset of the full KiDS DR4 catalog, optimized for weak lensing measurements through multiple quality criteria including detection significance thresholds, star-galaxy separation, shape measurement quality flags from \lensfit, and removal of sources near bright stars or survey boundaries \citep{giblin2021}. 
As explained in \autoref{sec:data}, we obtain the KiDS-1000 binary flag by cross-matching the DR4 data with the publicly released KiDS-1000 catalog, since the flag cannot be reproduced from the information available in the DR4 catalogs. 
Thus, we implement a machine learning approach that emulates the KiDS-1000 selection: a random forest classifier that distinguishes between galaxies included in KiDS-1000 versus those in the parent DR4 catalog. 
When doing inference, rather than applying a deterministic classification threshold, the predicted class is drawn from a Bernoulli distribution with probability equal to the predicted class probability given by the random forest.
Details of training and architecture are provided in \autoref{sec:selection_architecture}.
The classifier operates on 20 photometric features: \auto\ magnitude, \gaap\ magnitudes in all 9 bands ($ugriZYJHK_{\rm s}$), and limiting magnitudes derived from the 1-$\sigma$ flux uncertainties. 

The trained random forest classifier provides a practical approximation of the KiDS-1000 selection function. On the test set it reaches an overall accuracy of $\sim 63\%$ in a highly imbalanced regime where only $\sim 25$–$30\%$ of DR4 galaxies belong to KiDS-1000. A trivial classifier that labels all objects as non–KiDS-1000 would already achieve $\sim 70\%$ accuracy but zero recall; our performance therefore reflects a non‑trivial compromise between recovering KiDS‑1000 sources and limiting contamination. More informatively, the Receiver Operating Characteristic (ROC) curve, which plots the true positive rate against the false positive rate as the decision threshold is varied, has an area under the curve of $\simeq 0.90$. This quantifies the probability that the classifier assigns a higher score to a randomly chosen positive example than to a randomly chosen negative example, indicating strong intrinsic discrimination. At our fiducial operating point (defined by binomial sampling from the classifier's raw probability predictions) the classifier attains a precision of $\sim 0.44$ and a recall of $\sim 0.46$ for the KiDS‑1000 class; alternative thresholds could increase recall at the expense of purity.

Feature‑importance analysis shows that the $Z$‑ and $i$‑band \gaap\ magnitudes dominate, followed by the $r$‑band and near‑infrared \gaap\ magnitudes and the limiting magnitudes. The classifier maintains an accuracy close to the global value across well‑populated magnitude bins, with best performance around $r_\auto\simeq 21$--$23$, near the peak of the KiDS‑1000 number counts.

\autoref{fig:classifier_performance} further examines classifier behaviour and includes the weight predictions described in the next section. The weight distributions for galaxies classified as KiDS‑1000 members show the expected bimodal \lensfit structure, with true positives having systematically higher weights than false positives, confirming that the classifier preferentially selects galaxies with better shape‑measurement quality. At the ensemble level, the redshift distributions of selected galaxies closely match those of the true KiDS‑1000 sample in all tomographic bins, with differences in the mean redshift below the level relevant for our cosmological analysis.

\subsection{\lensfit weight prediction}
\label{sec:weight-prediction}

\begin{figure}
    \centering
    \includegraphics[width=0.99\linewidth, trim={0 11cm 0 1.1cm},clip]{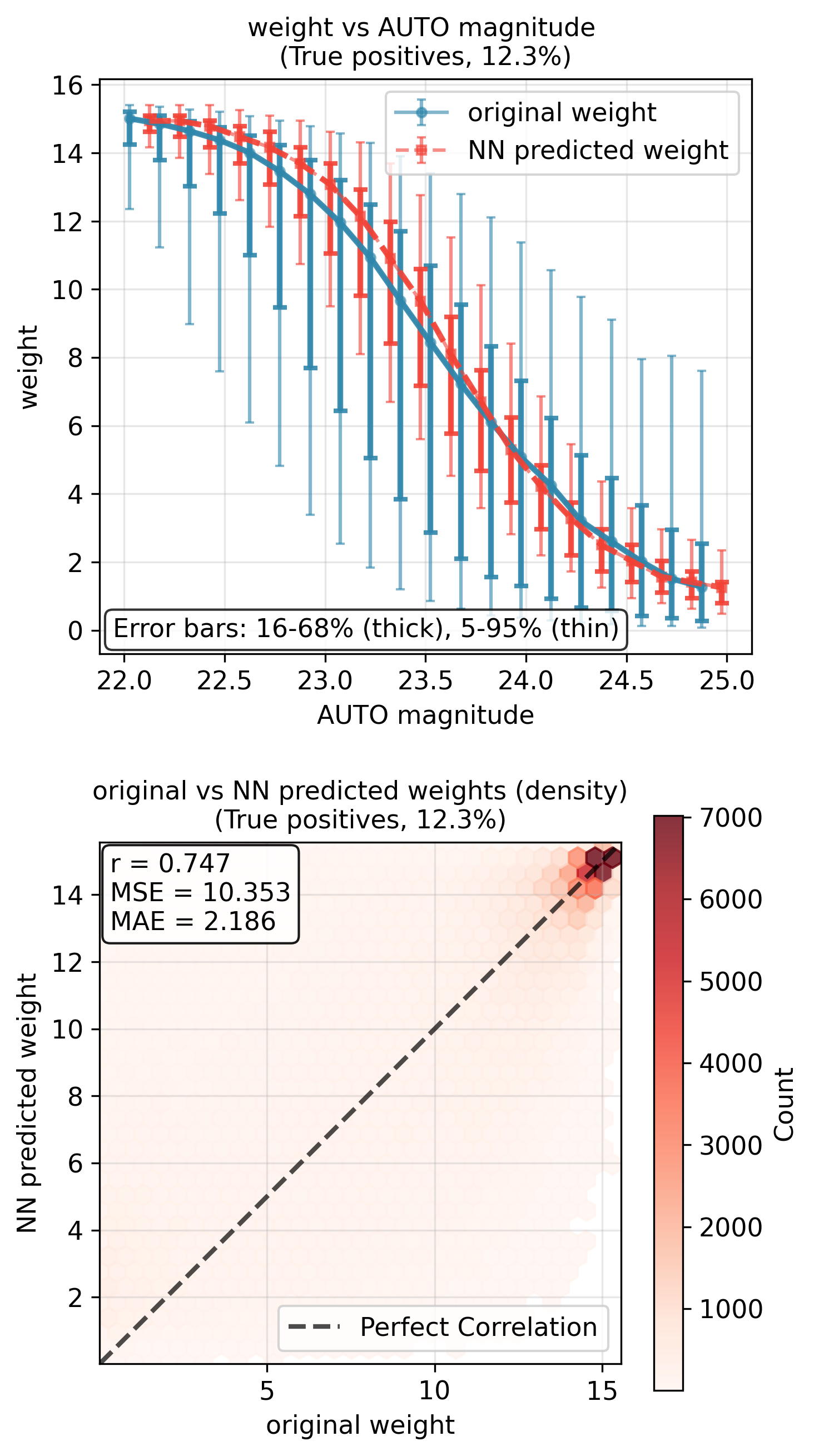}
    \caption{Performance of the neural network regressor for \lensfit weight prediction. Mean predicted weights (red) versus true weights (blue) as a function of \auto\ magnitude, showing good agreement despite underestimated scatter. }
    \label{fig:regressor_performance}
\end{figure}

The \lensfit weights provide a quantitative measure of shape‑measurement quality for weak lensing analyses \citep{miller2013}: they are inversely related to the variance of the shear estimates, so galaxies with better‑constrained shapes receive higher weights and contribute more reliably to the cosmic shear signal \citep{giblin2021}.
We implement a neural network regressor to predict \lensfit weights, from the same 20 photometric features as above. 
Details of training and architecture are provided in \autoref{sec:weight_architecture}.
Training data is restricted to KiDS-1000 galaxies only.
The trained neural network achieves good performance in predicting \lensfit weights. On the KiDS-1000 test sample the model attains a mean absolute error of $2.30$, a root mean squared error of $3.25$, and an $R^2$ score of $0.58$. This level of scatter is consistent with the fact that \lensfit weights depend on quantities that are not available in the catalog photometry, such as detailed galaxy morphology, local PSF quality, and proximity to bright stars.

As shown in \autoref{fig:regressor_performance}, the performance varies with magnitude. The model successfully reproduces the expected trend of decreasing weight with increasing $r_\auto$ and accurately captures the mean weight as a function of magnitude, with a typical bias of only $\sim 0.5$. The observed scatter in the true data is roughly twice as large as the model-predicted scatter, reflecting the missing non-photometric information, but this has negligible impact on the redshift distributions relevant for our analysis. The network recovers the overall shape of the weight distribution and performs particularly well in the critical high‑weight range of $10$--$15$. The residuals are approximately symmetric, with a mean of $-0.14$ and a standard deviation of $3.25$, indicating well‑behaved errors without strong systematic trends.

\section{Results}
\label{sec:results}

We now describe the synthetic KiDS DR4 and KiDS-1000 galaxy catalogs, obtained by drawing galaxies from one of the population models (\shark or \pc), and applying the data model (see \autoref{sec:data-model}). 
Finally, the KiDS-1000 selection (see \autoref{sec:k1k-selection}) and tomographic binning (see \autoref{sec:tomo-selection}) steps are applied in order to obtain mock galaxy samples in tomographic redshift bins.
\lensfit weights are derived with our emulator (see \autoref{sec:weight-prediction}).

\subsection{SKiLLS footprint}
\label{sec:results-imsims}

\begin{figure*}
    \centering
    \includegraphics[width=1\linewidth]{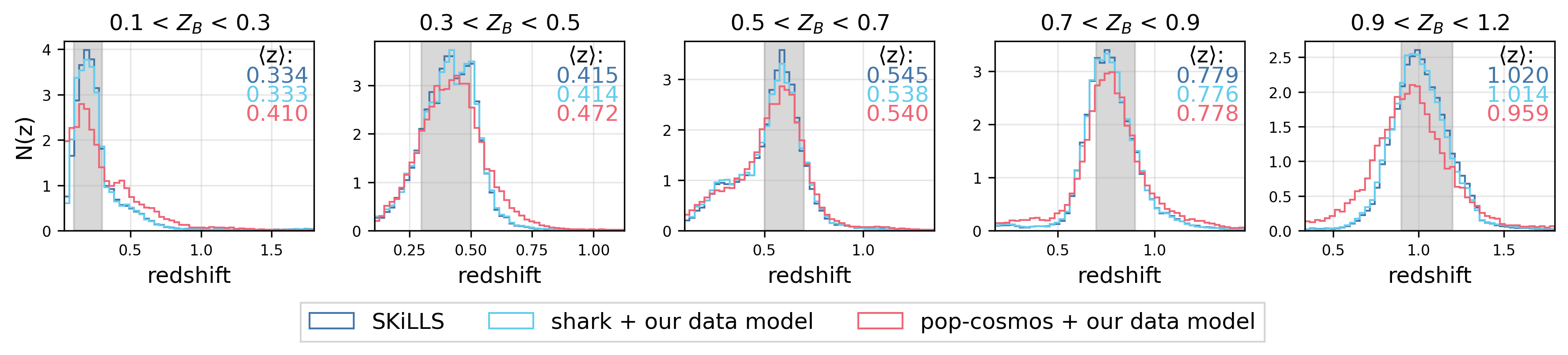}
    \caption{Redshift distributions in SKiLLS image simulations (dark blue), and from the \shark (light blue) and \pc (red) population models processed through our data model, all with \lensfit weights applied, shown in BPZ tomographic bins (titles). The grey band in each panel marks the BPZ $Z_B$ range defining the bin. The distinct $n(z)$ recovered by \pc\ and \shark are expected given the differences in their underlying modeled color–redshift relations (see text for details).}
    \label{fig:nofzs_imsims}
\end{figure*}

\autoref{fig:nofzs_imsims} shows the predicted redshift distributions of galaxies obtained by simulating the KiDS-1000 tomographic samples with the \shark and \pc population models, both interfaced with our data model.
Galaxies are injected at the depths (magnitude limits in the 9 bands) of objects from the SKiLLS footprint, so that the resulting galaxy samples have the same observing conditions as SKiLLS. 

First and foremost, this figure serves as a validation of our data model (Section~\ref{sec:data-model}): the redshift distributions obtained by combining \shark galaxies with our data model (light blue) are essentially identical to the `true' distributions from SKiLLS (dark blue). The differences between the means of the redshift distributions are all $< 0.005$, which is well within requirements for ongoing weak lensing surveys.
Given the large number of galaxies in each tomographic bin, the statistical uncertainties on the inferred $n(z)$ are negligible.

\begin{figure*}
    \centering
    \includegraphics[width=0.98\textwidth]{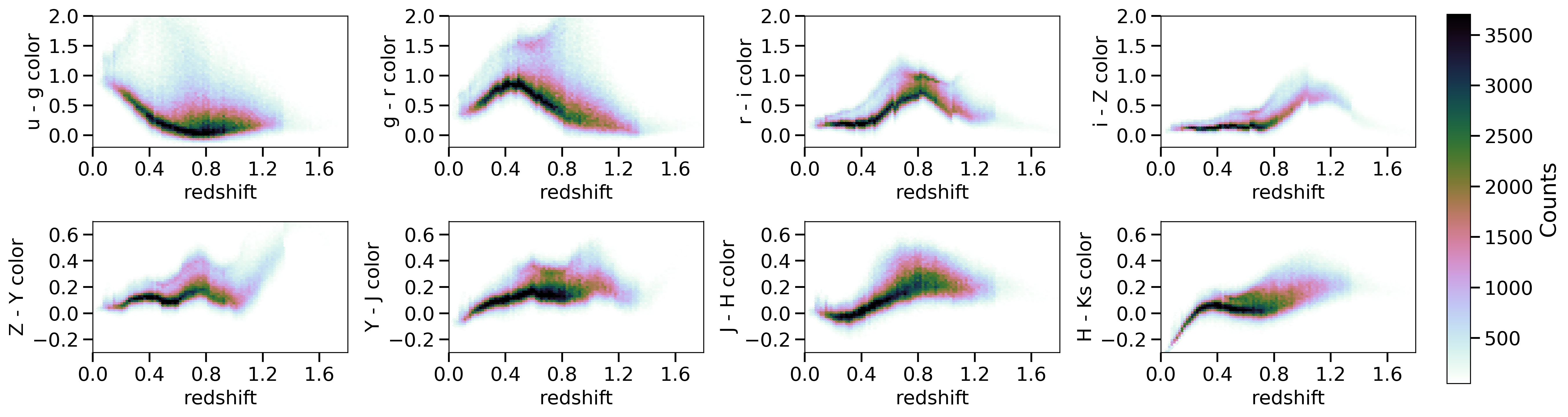}
    \caption{
    Noiseless color--redshift distributions for the \shark-based mock KiDS-1000 sample in the SKILLS footprint.
    }
    \label{fig:noiseless_colors_shark}
\end{figure*}

\begin{figure*}
    \centering
    \includegraphics[width=0.98\textwidth]{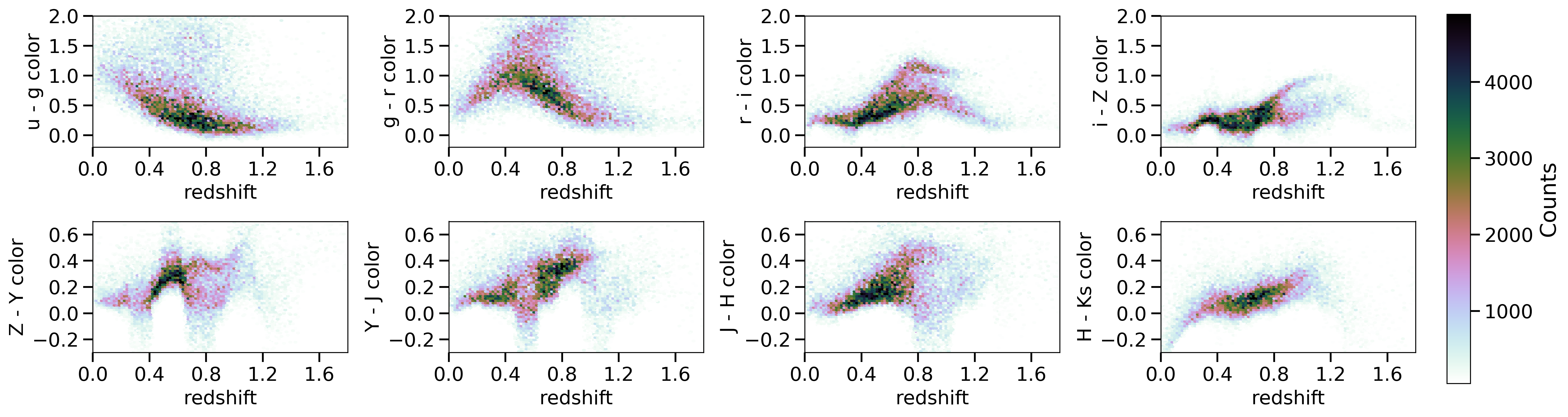}
    \caption{
    Noiseless color--redshift distributions for the \pc-based mock KiDS-1000 sample in the SKILLS footprint.
    }
    \label{fig:noiseless_colors_pop}
\end{figure*}

Replacing \shark for \pc mock galaxies (shown in red) results in more significant changes in the redshift distributions, with the largest shift in the highest-redshift bin ($\Delta z \sim 0.05$–$0.1$). 
This is the result of differences in the color--redshift relation between \shark, shown in \autoref{fig:noiseless_colors_shark}, and \pc, shown in \autoref{fig:noiseless_colors_pop}: at high redshifts, \pc galaxies occupy a broader range of generally redder colors than \shark (e.g.\ in $r-i$ and $i-Z$).
As a result, BPZ more often assigns intermediate-redshift \pc galaxies to the highest bin and scatters some high-redshift objects to lower $Z_B$, whereas \shark yields a simpler mapping of high-redshift colors into that bin.
However, we cannot link these color differences to discrepancies in physical galaxy properties  since no physical parameters are available for \shark.
These differences exemplify the contrasting philosophies of the two approaches: \pc's data-driven approach directly learns the full diversity of observed color--redshift correlations, including population scatter; \shark's physics-based prescriptions produce more uniform color evolution that successfully matches aggregate properties such as luminosity functions and stellar masses, but not the detailed color--redshift distributions.

The ridges and bifurcations visible in the \pc color--redshift relation, particularly in the infrared bands, arise from strong nebular emission lines (H$\alpha$, [O\,{\sc{iii}}], H$\beta$) transitioning across filter bandpasses. 
This likely reflects differences in emission line treatment: \pc\ includes more detailed nebular physics \citep{leistedt2023hierarchical, thorp2025}, higher equivalent widths for star-forming galaxies, and greater diversity in star formation histories, while \shark's smoother tracks suggest weaker emission line contributions or simplified prescriptions. 
Star-forming galaxies with high specific star formation rates (sSFR) show the strongest features due to larger equivalent widths, while quiescent populations follow smoother continuum-dominated tracks. 
These differences in the color--redshift relation affect the $n(z)$ via the tomographic bin assignments, performed with BPZ $Z_B$ \footnote{Photo-$z$ codes rely on capturing sharp color features to break degeneracies and assign accurate redshifts.}.

\subsection{KiDS DR4 data footprint}
\label{sec:results-data}

\begin{figure*}
    \centering
    \includegraphics[width=1\linewidth]{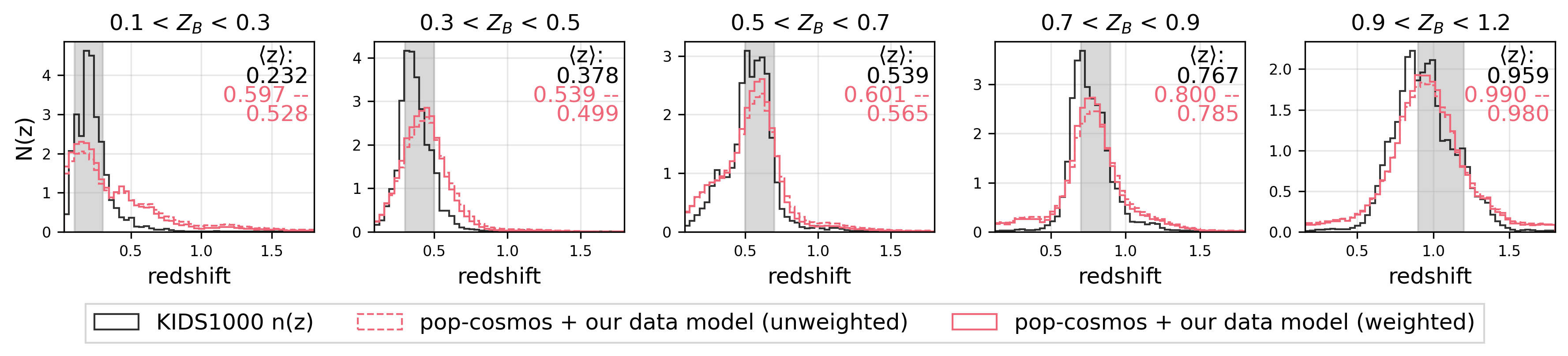}
    \caption{Redshift distributions in KiDS-1000 (black) and in the \pc\ population model processed through our data model. Solid red histograms include \lensfit weights, while dashed red histograms are unweighted. The grey band in each panel marks the BPZ $Z_B$ range defining the bin. There are inherent differences between the samples compared here (see text).}   
    \label{fig:nofzs_data}
\end{figure*}

We now simulate galaxies at the depths of the real DR4 data, in order to span the same range of observing conditions.
We pick 200 random tiles, which provides a sample representative of the DR4 data footprint.
For each real KiDS-1000 object in these tiles, we draw from our model until a simulated object is accepted  by our KiDS-1000 selection model.
The resulting redshift distributions are shown in \autoref{fig:nofzs_data}.
The differences with \shark (not shown) are similar to those shown in \autoref{fig:nofzs_imsims}, which is expected since the depth distributions of KiDS DR4 and SKiLLS are so closely comparable.

We compare our results to the publicly released KiDS-1000 redshift distributions \citep{Hildebrandt2021}. These were obtained by re-weighting deep spectroscopic reference catalogues, including zCOSMOS \citep{Lilly2007, Lilly2009}, VVDS-Deep \citep{LeFevre2005, LeFevre2013, LeFevre2015}, and DEEP2 \citep{Newman2013}, using a self-organizing map \citep[SOM;][]{Kohonen1982, Masters2015} to match the nine-dimensional magnitude-space of the source galaxies \citep{Wright2020a, Hildebrandt2021}, with sources limited to a `gold' selection that only includes regions of magnitude-space covered by the reference samples \citep{Wright2020a}. 
In that framework, the accuracy of the calibration is limited by the depth, area and selection of the spectra, by cosmic variance in the calibration fields, and by the exclusion of sources in poorly sampled regions of color–magnitude space.

By contrast, our forward-modeling calibration is not tied to a particular spectroscopic training set. 
The differences between the publicly released redshift distributions could be due to differences in the BPZ configuration between DR4 and DR5 (leading to differences in tomographic binning), or differences in the `gold' selection, and, most importantly, from
the fact that the spectroscopic reference sample used in the SOM approach does not necessarily span the same galaxy population as captured by the \texttt{pop-cosmos} model at KiDS depth.
 For reference, \citet{Hildebrandt2021} and \citet{Wright2025} quote mean-redshift uncertainties of $\sigma_{\langle z \rangle} \simeq 0.01$--$0.05$ at 95\% confidence per tomographic bin for the SOM-based calibration. As emphasized previously, the underlying galaxy samples compared in \autoref{fig:nofzs_data} are somewhat different. Hence any systematic shifts in cosmological parameters between the two approaches can only be assessed in a full cosmological analysis, which we will present in an upcoming paper.

\subsection{Galaxy populations in tomographic bins}
\label{subsec:mass_sSFR_bins}

\begin{figure*}
    \centering
    \includegraphics[width=0.98\textwidth]{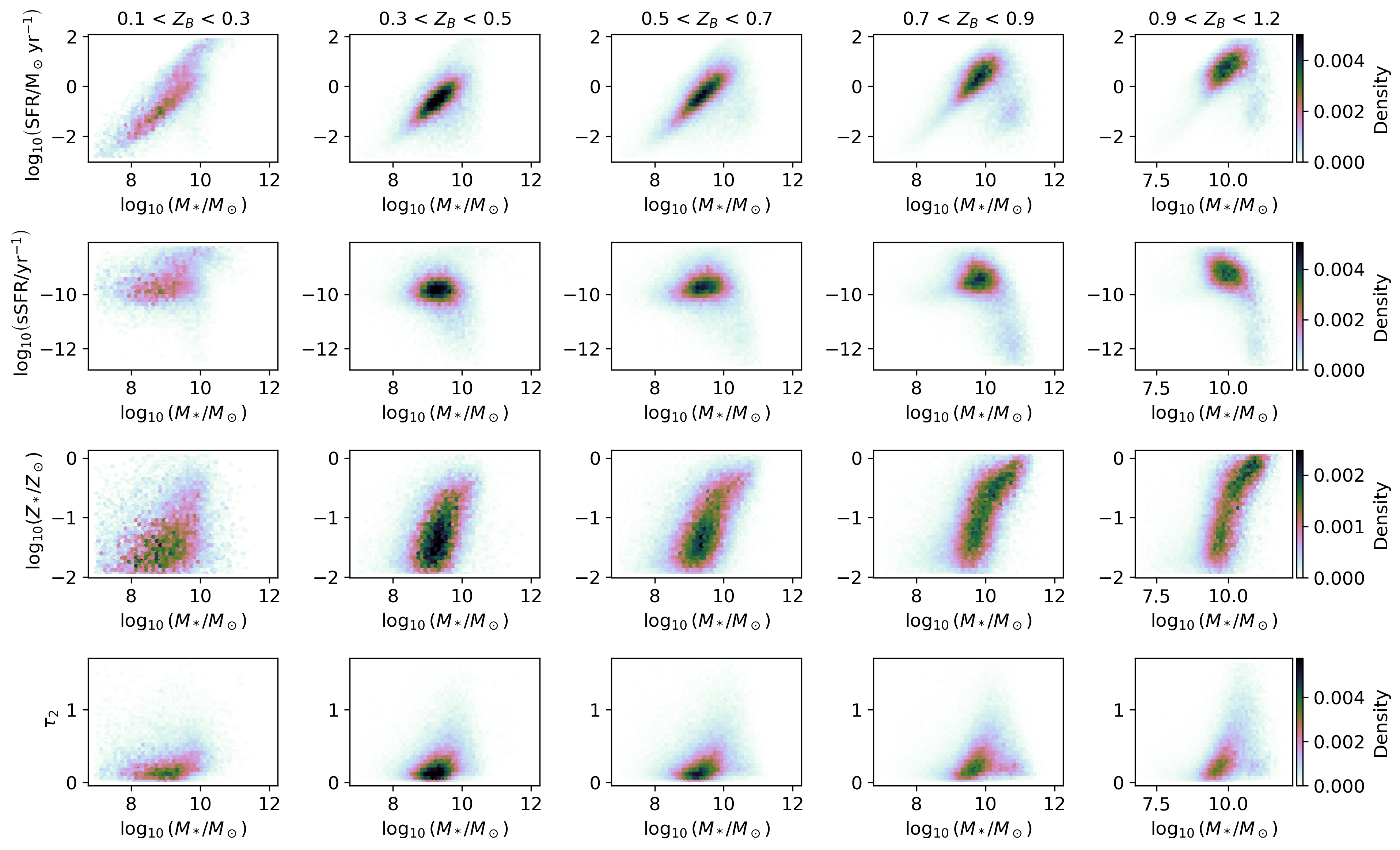}
    \caption{
    Stellar mass and other properties of \pc\ galaxies in the KiDS-1000 tomographic bins. Panels show the joint distribution of stellar mass with SFR (top row), sSFR (second row), stellar metallicity (third row), and diffuse dust optical depth (bottom row), for galaxies assigned to a given BPZ tomographic redshift bin (columns).
    }
    \label{fig:mass_sfr_tomobins}
\end{figure*}

The intrinsic properties of \pc galaxies in the simulated KiDS-1000 sample are illustrated in \autoref{fig:mass_sfr_tomobins}. 
Each row shows a different physical parameter as a function of stellar mass ($M_\star/M_\odot$): 
star formation rate (SFR); specific SFR (sSFR); 
stellar metallicity, $Z_\star/Z_\odot$; and diffuse dust optical depth, $\tau_2$. The first two rows highlight the star-forming main sequence, along which mass and SFR correlate closely (\citealp{noeske2007, daddi2007}; for a literature survey, see e.g.\ \citealp{speagle2014, popesso2023}), as well as the transition at high stellar mass between the star-forming and passive sequences \citep[e.g.][]{kauffmann2003, baldry2006, bower2017}. The third row probes the stellar mass--metallicity relation \citep[e.g.][]{gallazzi2005, gallazzi2014, zahid2017}, whilst the fourth highlights the correlation between dust attenuation and stellar mass \citep[e.g.][]{garn2010, zahid2013, zahid2017}.

The plotted quantities are the noiseless SPS parameters of the \pc galaxies that both pass the KiDS-1000 selection and are assigned to the corresponding tomographic bin based on their BPZ redshift. 
This reveals how the KiDS-1000 selection function and detection thresholds shape the distribution of intrinsic galaxy properties. 
At low redshift (first column), the sample spans $10^7 \lesssim M_\star/\mathrm{M}_\odot \lesssim 10^{11}$, including both star-forming galaxies on the main sequence and massive quiescent systems ($\mathrm{sSFR} \lesssim 10^{-11}\,\mathrm{yr}^{-1}$). 
The expected scaling relations emerge naturally: more massive galaxies are older, more metal-rich, and dustier, particularly on the star-forming sequence.
With increasing redshift, the mass distribution shifts toward $M_\star \gtrsim 10^{10}\,\mathrm{M}_\odot$ due to the detection limit (effectively Malmquist bias), while the star-forming main sequence evolves to higher SFR and sSFR at fixed mass.
In the higher redshift bins (fourth and fifth columns), clear bimodality appears between star-forming galaxies ($\mathrm{sSFR} \simeq 10^{-9}\,\mathrm{yr}^{-1}$) and massive quiescent systems.
At fixed mass, higher-redshift galaxies are younger, slightly less metal-rich, and show increased dust optical depth with both mass and SFR.
These correlated trends in mass, SFR, metallicity, and dust shape the color--magnitude--redshift distribution that drives photometric redshift estimation and tomographic calibration.

It is instructive to compare \autoref{fig:mass_sfr_tomobins} with Figure~12 of H26, where similar relations between stellar mass and inferred physical properties are shown for the real KiDS-1000 galaxies, and with tomographic bins defined using \pc-based photometric redshifts rather than the BPZ $Z_B$. Qualitatively, the same structures are visible in both figures: a well-defined star-forming main sequence and a population of massive quenched galaxies in each tomographic bin, together with increasing characteristic stellar mass and sSFR toward higher redshift, and tightening mass--metallicity relations. 
However, H26 finds slightly sharper and more coherent sequences within each bin, with less leakage of low- or high-redshift populations across bin boundaries, while our BPZ-based bins display somewhat broader and more blended structures, especially at the edges of the tomographic ranges. 
This contrast is consistent with the higher redshift accuracy of \pc relative to BPZ, and it illustrates explicitly how improved photo-$z$ performance tightens the mapping between physical properties and tomographic bin.

\begin{figure*}
    \centering
        \includegraphics[width=0.98\textwidth]{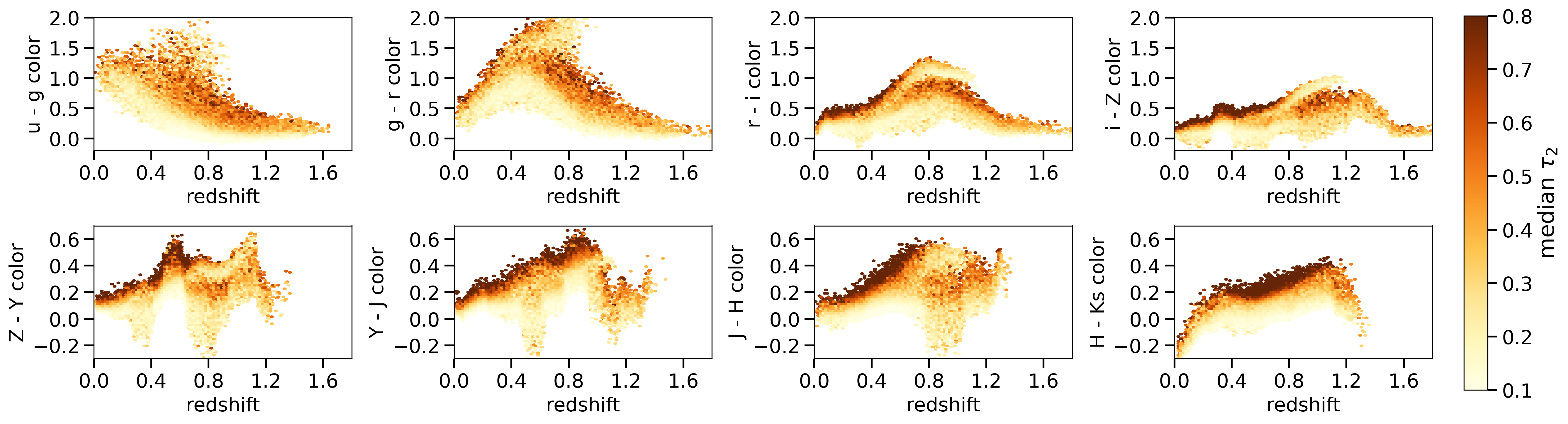}
        \label{fig:colors_vs_z_dust2}
        \includegraphics[width=0.98\textwidth]{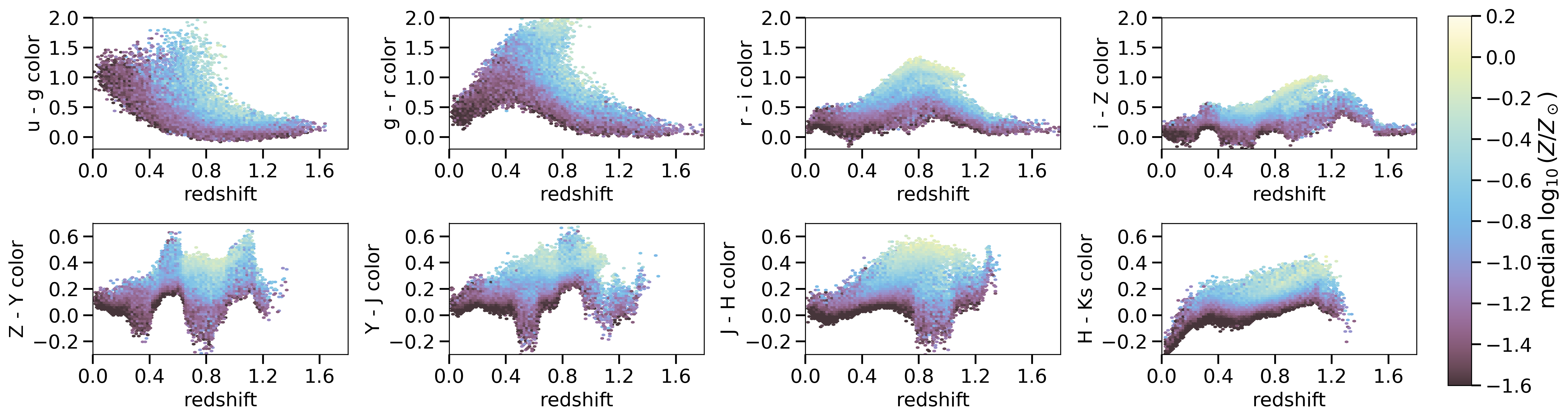}
        \label{fig:colors_vs_z_log10Z}
        \includegraphics[width=0.98\textwidth]{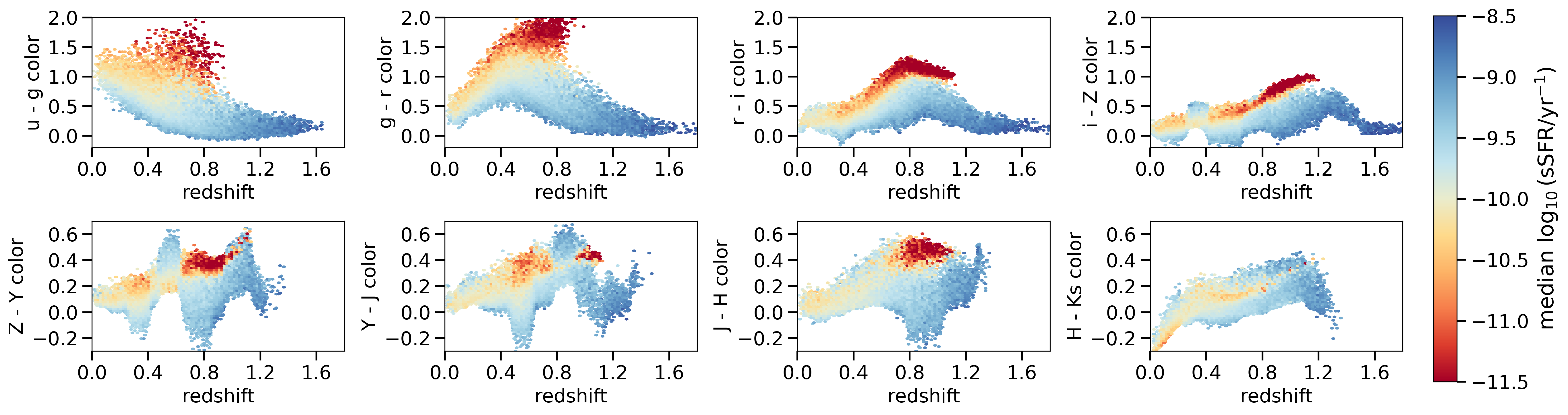}
        \label{fig:colors_vs_z_sSFR}
 \includegraphics[width=0.98\textwidth]{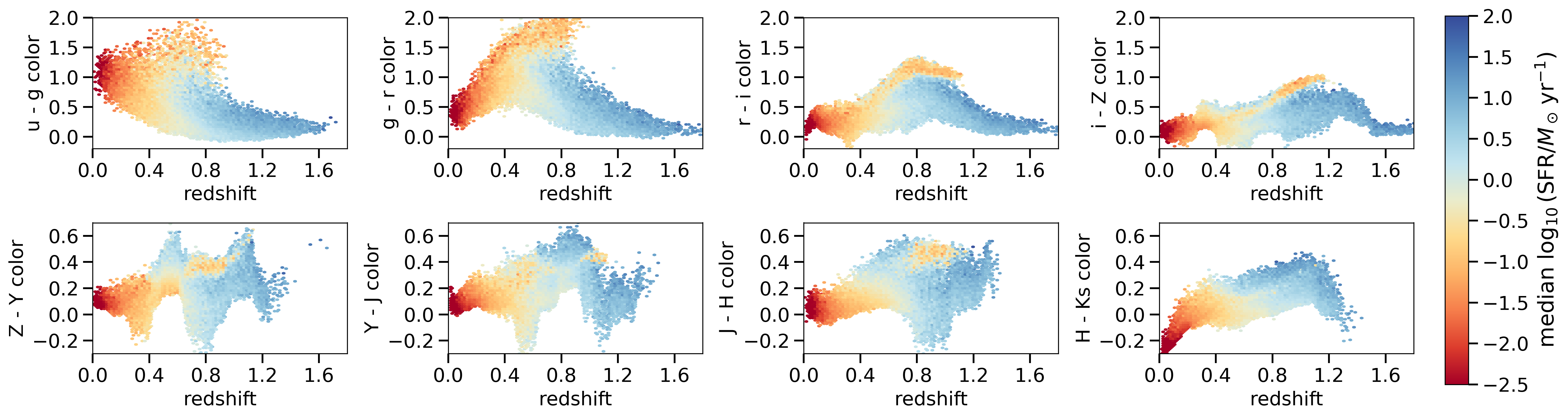}
        \label{fig:colors_vs_z_SFR}
        \includegraphics[width=0.98\textwidth]{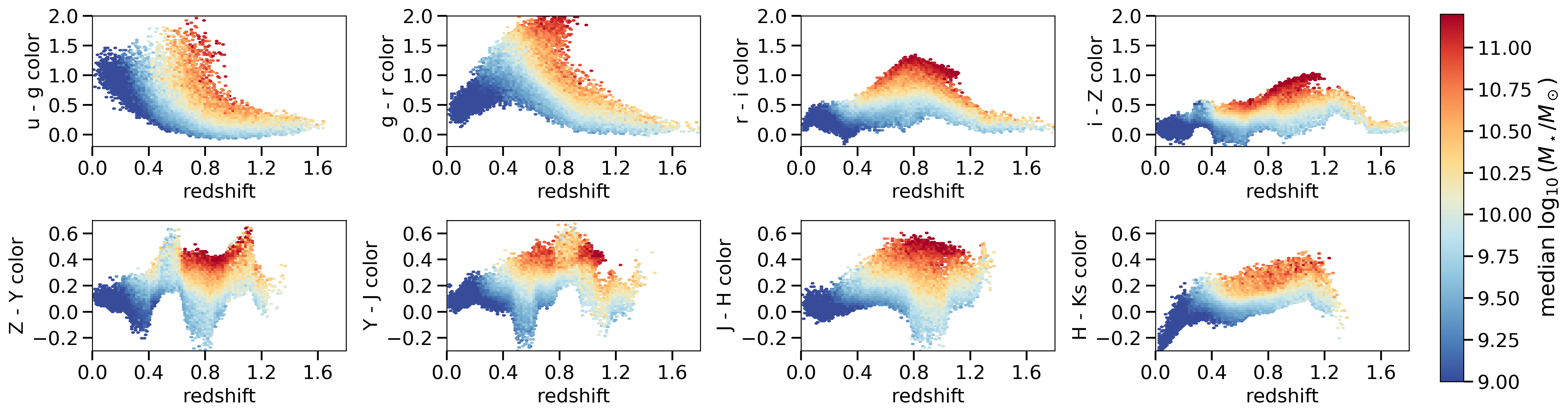}
        \label{fig:colors_vs_z_M_formed}

    \caption{
    Noiseless color--redshift relations colored by physical properties of the \pc-based  KiDS-1000 mocks. 
    }
        \label{fig:noiseless_colors_pop_properties}
\end{figure*}

In addition to the relations between stellar mass and other properties, it is useful to examine how these same populations populate color--redshift space. 
This is shown in \autoref{fig:noiseless_colors_pop_properties}.
These maps reveal the familiar sharp features associated with strong emission lines moving through the filters, as well as smooth trends whereby redder optical and NIR colors at fixed redshift correspond to higher mass, higher metallicities, larger dust optical depths, and lower sSFRs. 
Conversely, the bluest regions of each color–redshift relation are populated by low-mass, young, metal-poor galaxies with high specific star-formation rates.
This can be compared directly to Figures~7--11 of H26, in which analogous color–redshift relations are plotted with the physical properties (including redshift) inferred from full SED fitting of individual KiDS-1000 galaxies. 
Qualitatively, the same structures are visible in both figures: the positions and shapes of the emission-line-induced breaks in color with redshift, the monotonic gradients of dust, metallicity and sSFR across the color--redshift plane, and the clear separation between star-forming and quenched populations. 
The main difference (causing the increased horizontal blurring of Figures.~7--11 of H26) is the result of the horizontal axis being the inferred photometric estimate as opposed to the true model redshift used here.

\subsection{Number counts in tomographic bins}
\label{sec:number_counts}

The numbers of galaxies in each tomographic bin can differ by $\sim 5$-$10\%$ between our mock catalogs and the SKiLLS image simulations or the KiDS data.
The reasons for these differences are the changes in BPZ configuration outlined in \autoref{sec:tomo-selection}, the differences in 9-band depths in the SKiLLS and DR4 data footprints, and the KiDS-1000 classifier employed.
We isolated these effects and found that the tile-to-tile changes in 9-band depth combined with the choice of training data of the classifier is the most important effect.
This is because the KiDS‑1000 classifier effectively imposes a magnitude- and quality-dependent selection. 
Small tile-to-tile changes in limiting magnitudes shift the effective flux threshold, strongly affecting how many galaxies pass the cut near the survey flux limit. 
When the classifier is trained on SKiLLS or on the DR4 data, it sees different depth distributions and flag definitions, so it learns slightly different decision boundaries. 
Together, these effects change the selection efficiency and thus the number of galaxies in each tomographic bin, especially when training the classifier on the data and applying it to SKiLLS, or conversely.
Nevertheless, we find that the shapes of the redshift distributions in tomographic bins are robust to these changes.

\section{Discussion}
\label{sec:discussion}

The comparison between \shark and \pc underscores the advantage of an empirically calibrated population model.

The \pc model has been extensively validated through several companion studies \citep{alsing2024, thorp2024, thorp2025, deger2025}. 
These works demonstrate that it reproduces expected color--redshift relations and galaxy evolution trends, with per-galaxy photometric redshifts achieving scatter and outlier rates competitive with or better than other available methods \citep{thorp2024}. 
In a companion paper (H26), the updated population model of \citet{thorp2025} are applied to individual KiDS DR4 galaxies, yielding accurate photometric redshifts and physical property estimates for individual objects. 
Together, these studies establish \pc as empirically well-calibrated in the optical and NIR at KiDS depths. 

\shark provides a complementary perspective through its semi-analytic approach, which we use to probe systematic shifts under different galaxy‑formation assumptions.
Existing \shark validation studies \citep{lagos2018, Lagos_2020, Bravo_2020} have successfully tested predictions against luminosity functions across cosmic time, $z\simeq 0$ color distributions, submillimeter number counts, and color–color diagrams at $z\simeq 2$-$3$.
However, these validations focus on integrated properties and specific redshift slices rather than color evolution as a continuous function of redshift at $0.5 \lesssim z \lesssim 2$.
Figure 7 of H26 reveals that the color bimodality at $z>0.4$ due to the separation between the red sequence and the blue cloud is better reproduced by \pc (\autoref{fig:noiseless_colors_pop}) than \shark (\autoref{fig:noiseless_colors_shark}). 
The systematic shifts we observe ($\Delta z \sim 0.05$--$0.1$) when substituting them therefore highlight the sensitivity of forward-modeled redshift distributions to the calibration of the underlying population model.

The current implementation of our forward modeling framework presented here does, inevitably, have some limitations.  However, our tests show no significant impact on $n(z)$ calibration or scientific conclusions. We now discuss these points in turn:

\begin{itemize}

\item
The \pc model does not currently include galaxy morphological parameters (size, ellipticity, Sérsic index), which are known to correlate with photometric properties. 

To investigate this correlation and any potential systematic shifts in the magnitude and color distributions, we have made tests with HSC-SSP data\footnote{The information to run such tests is not available for KIDS.} \citep{Aihara2022_HSC_PDR3}, where KiDS-like flux and image-level cuts were applied to HSC shape catalogues. We find that such morphology- and resolution-dependent cuts mainly affect the extreme tails of the color distributions and leave their bulk largely unchanged (Tudorache et al.~in prep.). 
This demonstrates that the absence of explicit morphology in our forward modeling has a negligible impact on the color–redshift relation and hence on the tomographic $n(z)$ calibration.

\item
Our data model is trained exclusively on the SKiLLS image simulations, which have their own limitations.

The SKiLLS noise model uses a single-image approximation in the infrared bands, rather than fully simulating the VIKING paw-print observing strategy \citep{edge2013}, and in the optical bands assumes Gaussian noise with empirically-calibrated variance. 
While these approximations produce realistic photometric scatter, they may not capture all subtleties of the noise properties, such as spatial correlations or non-Gaussian tails. 

Additionally, the SKiLLS PSF modeling uses position-dependent polynomial reconstructions for the $r$-band and Moffat profiles for other bands \citep{miller2013}, which may introduce small systematic differences compared to the actual survey PSFs. 

\end{itemize}

These limitations are unlikely to significantly impact our redshift distribution estimates, as we condition our models on depth-dependent quantities and train on realistic detected galaxies.
However, as surveys push toward higher precision on cosmological parameters,  it will become essential to propagate residual sources of uncertainty in redshift calibration.

In principle, one could imagine calibrating $n(z)$ by combining the redshift posterior distributions from full SED inference on every source galaxy \citep{Leistedt2016, malz22, autenrieth2024}. However, this is not practical for KiDS-1000: with only nine broad bands and many sources near the detection limits, individual-galaxy posteriors remain broad and sometimes multi-modal, even under the \pc prior (H26). This would propagate large per-object scatter into the $n(z)$. Moreover, such an approach would still need to model the full survey selection function. Our forward-modelling framework does this explicitly by propagating the population prior through the data model, making it a more robust route to $n(z)$ calibration.

After our DR4 data preparation pipeline was finalized KiDS Data Release 5 \citep{wright2024} was made available.
Differences with DR4 include increased sky coverage, additional $i$-band observations, and improved astrometry, photometry, and redshift calibration \citep{Wright2025}.
However, the KiDS-1000 weak lensing catalog \citep{giblin2021, Hildebrandt2021} that we model in this work is a DR4 data product, with the equivalent DR5 weak lensing analysis having only been published very recently \citep{wright2025legacy}.

\section{Conclusions}
\label{sec:conclusions}

We have developed a forward-modeling approach that infers galaxy redshift distributions directly from a generative galaxy population model combined with a realistic survey data model. Specifically, we combined the empirically calibrated \pc population model with a data model learned from KiDS image simulations \citep[SKiLLS;][]{Li_2023} to forward-model the KiDS-1000 weak-lensing sample.

Our key findings are:

\begin{itemize}
    \item \textbf{Data-model validation:} Machine learning models can accurately capture the complex transformations between intrinsic galaxy properties and observed photometry in KiDS, successfully reproducing the statistical properties of the SKiLLS synthetic galaxy catalogs when using \shark, the galaxy population that was used to generate these simulations. 
    \item \textbf{Importance of empirically calibrated population model:} Replacing \shark by the \pc galaxy-population model leads to shifts of $\Delta z \simeq 0.05$-$0.1$ in the first and last tomographic bins, demonstrating that population assumptions can substantially influence the predicted tomographic redshifts. These differences arise primarily from variations in the galaxy color–redshift relations near the tomographic bin boundaries, underscoring that careful empirical calibration of the galaxy population model is essential for robust redshift calibration.
    \item \noindent\textbf{Generalisation of \pc beyond its training data:} The forward-modeled KiDS-1000 redshift distributions obtained demonstrate that the COSMOS2020-trained \pc model can be applied successfully to a survey with different depth, selection, and observing conditions. This conclusion is further corroborated by the companion analysis H26, in which MCMC inference under the \pc prior is applied to KiDS-1000 galaxies cross-matched with DESI DR1 Bright Galaxy and Luminous Red Galaxy samples, yielding accurate photometric redshifts with low outlier fractions that satisfy Stage~IV weak-lensing requirements. 
    \item \textbf{Redshift distribution estimation:} The framework presented here represents a practical tool for estimating redshift distributions and quantifying  systematic uncertainties. By combining the realism of data-driven population models with the efficiency of machine-learning-based observational models, we can efficiently explore the impact of different assumptions about galaxy populations on cosmological measurements. 
\end{itemize}

Building on the method developed here -- and on the companion analysis of H26 -- we plan to perform a full KiDS cosmological analysis using \pc-based redshift distributions. 
As Stage~IV surveys push toward percent-level precision on cosmological parameters, such forward-modeling tools, grounded in empirically-calibrated population models such as \pc\ will be essential to ensure that systematic uncertainties in redshift calibration do not limit our ability to constrain fundamental physics.

\section*{Data Availability}
We have used the gold sample of weak lensing and photometric redshift measurements from the fourth data release of KiDS \citep{kuijken2019, Wright2020a, Hildebrandt2021, giblin2021}. Cosmological parameter constraints from KiDS-1000 have been presented in \citet{asgari2021}, \citet{heymans2021} and \citet{troster2021}, with the methodology presented in \citet{joachimi2021}. 
The SKiLLS simulations from \citet{Li_2023} are available on \texttt{SurfDrive}\footnote{\url{https://surfdrive.surf.nl/files/index.php/s/iSvDmHQJjDa0ewG}}, and are based on input galaxies from SURFS\footnote{\url{https://surfdrive.surf.nl/files/index.php/s/uegK5tc15TbWib7}} \citep{elahi2018} and stars from TRILEGAL\footnote{\url{https://surfdrive.surf.nl/files/index.php/s/dMsqnkeEUFSSLHE}} \citep{girardi2005}.

\section*{Author Contributions}
We outline the different contributions below using keywords based on the Contribution Roles Taxonomy (CRediT; \citealp{brand15}). \textbf{BL:} 
conceptualization; 
methodology; 
software; 
formal analysis; 
investigation; 
data curation; 
visualization; writing -- original draft; 
writing -- review \& editing.
\textbf{HVP:} 
conceptualization;
methodology;
visualization;
investigation;
validation;
writing -- review \& editing;
supervision; 
project administration; 
funding acquisition.
\textbf{AH:}
investigation;
software;
validation;
visualization;
writing -- review \& editing.
\textbf{ST:}
methodology;
software;
writing -- review \& editing. 
\textbf{DJM:}
conceptualization;
methodology;
investigation;
validation;
writing -- review \& editing.
\textbf{AL:}
conceptualization; 
data curation;
formal analysis;
investigation;
methodology;
software;
validation;
visualization.
\textbf{JA:}
conceptualization; 
data curation;
formal analysis;
investigation;
methodology;
software;
validation;
visualization.
\textbf{GJ:} 
software. 
\textbf{MNT:}
investigation;
validation;
writing -- review \& editing;
visualization.
\textbf{SD:}
investigation;
writing -- review \& editing.
\textbf{JL:}
writing -- review \& editing;
validation.
\textbf{BVdB:} 
validation; 
visualization.
\textbf{AHW:} 
data curation; 
writing - review \& editing.
\textbf{SSL:} 
resources;
writing - review \& editing.
\textbf{KK:} 
resources;
writing - review \& editing.
\textbf{HH:} 
data curation; 
writing - review \& editing.
\begin{acknowledgements}
BL is supported by the Royal Society through a University Research Fellowship. This work has been supported by funding from the European Research Council (ERC) under the European Union's Horizon 2020 research and innovation programmes (grant agreement no.\ 101018897 CosmicExplorer). This work has been enabled by support from the research project grant ‘Understanding the Dynamic Universe’ funded by the Knut and Alice Wallenberg Foundation under Dnr KAW 2018.0067. HVP was additionally supported by the G\"{o}ran Gustafsson Foundation for Research in Natural Sciences and Medicine.
AHW is supported by the Deutsches Zentrum für Luft- und Raumfahrt (DLR), under project 50QE2305, made possible by the Bundesministerium für Wirtschaft und Klimaschutz, and acknowledges funding from the German Science Foundation DFG, via the Collaborative Research Center SFB1491 ``Cosmic Interacting Matters - From Source to Signal''.
\end{acknowledgements}

\facilities{Based on observations made with ESO Telescopes at the La Silla Paranal Observatory under programme IDs 177.A-3016, 177.A-3017, 177.A-3018 and 179.A-2004, and on data products produced by the KiDS consortium. The KiDS production team acknowledges support from: Deutsche Forschungsgemeinschaft, ERC, NOVA and NWO-M grants; Target; the University of Padova, and the University Federico II (Naples).}

\software{
  \texttt{Astropy} \citep{astropy2013, astropy2018, astropy2022},
  \texttt{NumPy} \citep{harris2020},
  \texttt{SciPy} \citep{virtanen2020},
  \texttt{Matplotlib} \citep{hunter2007},
  \texttt{scikit-learn} \citep{scikit-learn},
  \texttt{GALSim} \citep{rowe2015},
  \texttt{FSPS} \citep{conroy2009, conroy2010, conroygunn2010},
  \texttt{Prospector} \citep{johnson2021},
  \texttt{CLOUDY} \citep{ferland2013},
  \texttt{Speculator} \citep{alsing2020},
  \texttt{pop-cosmos}\footnote{\url{https://github.com/Cosmo-Pop/pop-cosmos}} (\citealp{alsing2024, thorp2025}),
  \texttt{flowfusion}\footnote{\url{https://github.com/Cosmo-Pop/flowfusion}},
  \texttt{torchdiffeq} \citep{chen18},
  and the KiDS and SKiLLS reduction pipelines
  \citep{kuijken2019, Li_2023}.
}

%
%
\bibliographystyle{aasjournalv7}
\bibliography{spsbhm}

\appendix

\section{Machine Learning Architectures}

\subsection{Random forest classifier for detection}
\label{sec:detection_architecture}

The detection model employs a random forest classifier to predict whether galaxies are detected in the KiDS DR4 $r$-band observations. Random forests are ensemble learning methods \citep{breiman2001random, scikit-learn} that construct multiple decision trees during training and output the class that is the mode of the classes from individual trees. Our implementation uses 100 trees with entropy as the splitting criterion for information gain. The maximum tree depth is set to 37, with a minimum of 12 samples required for splitting and 32 samples per leaf node. The model operates on only two input features: the noiseless $r$-band magnitude and the \gaap\ limiting magnitude in the $r$-band. This simplified feature space reflects the assumption that detection is primarily determined by apparent magnitude and survey depth. The classifier was trained on 800,000 objects randomly sampled from the SKiLLS simulations, producing a binary classification output indicating whether each galaxy would be detected or not detected in the survey.

\subsection{Conditional flow for the uncertainty model}
\label{sec:uncertainty_architecture}

The uncertainty model predicts the \auto\ magnitude error conditioned on the \gaap\ magnitude error and noiseless magnitude, all in the $r$-band. This model employs a conditional ODE flow framework \citep{lipman2023flow,tong2023conditional}, which learns to transport samples from a simple base distribution to the target conditional distribution via continuous normalizing flows \citep{chen18, grathwohl18}. The architecture consists of a densely connected neural network with four hidden layers, each containing 128 neurons with hyperbolic tangent activation functions. Training proceeds by minimizing the flow matching objective \citep{lipman2023flow} using the \texttt{Adam} optimizer \citep{kingma14}. The conditional flow framework allows flexible modeling of complex conditional probability distributions without restrictive parametric assumptions about the error distribution, while enabling efficient sampling through deterministic ODE integration. All training data comes from detected galaxies in the SKiLLS image simulations, ensuring the learned distribution is conditioned on successful detection.

\subsection{Conditional ODE flow for the noise model}
\label{sec:noise_architecture}

The noise model generates correlated flux noise residuals across all nine photometric bands ($ugriZYJHK_{\rm s}$) plus the $r$-band \auto\ flux, conditioned on the noiseless model magnitudes and magnitude limits (equivalent to fluxes and flux uncertainties). The architecture mirrors that of the uncertainty model, using a conditional ODE flow framework \citep{lipman2023flow,tong2023conditional} with a densely connected neural network consisting of four hidden layers of 128 neurons each and hyperbolic tangent activation functions. However, the input dimensionality is substantially larger at 20 dimensions: nine noiseless magnitudes, ten flux uncertainties (\gaap\ for nine bands plus \auto\ for $r$-band), and the flow time parameter $t$. The output is ten-dimensional, providing predictions for the normalized flux residuals. 
The model is trained using the flow matching objective with the \texttt{Adam} optimizer, exclusively on detected SKiLLS galaxies to ensure proper conditioning on the detection process. By modeling the joint distribution of flux residuals across all bands simultaneously, the conditional flow framework naturally captures inter-band correlations without requiring explicit correlation matrices or parametric assumptions.

\subsection{Random forest classifier for KiDS-1000 selection}
\label{sec:selection_architecture}

This classifier predicts which galaxies from the parent DR4 catalog pass the quality cuts for inclusion in the KiDS-1000 weak lensing sample. The architecture employs 100 decision trees with entropy-based splitting, maximum tree depth of 37, minimum samples for split of 12, and minimum samples per leaf of 32. The input feature space is substantially richer than the detection model, comprising 20 features: \auto\ magnitude, and \gaap\ magnitudes in all nine bands, their corresponding magnitude uncertainties, and limiting magnitudes derived from the $1\sigma$ flux uncertainties. 
The training set consists of approximately 5 million galaxies split 70/30 for training and testing. During inference, we stochastically assign mock galaxies to the KiDS-1000 sample by drawing from a Bernoulli distribution using the classifier predicted probabilities.

\subsection{Neural network regressor for \lensfit weights}
\label{sec:weight_architecture}

This model predicts \lensfit shape measurement quality weights from photometric features alone, without access to morphological information or image-level diagnostics. The architecture consists of a fully-connected feed forward neural network with four hidden layers plus one output layer. Each hidden layer contains 128 neurons with hyperbolic tangent activation functions. The input features are identical to those used in the KiDS-1000 selection classifier: 20 photometric measurements including \gaap\ and \auto\ magnitudes, their uncertainties, and limiting magnitudes across all nine bands. Input features are standardized using training set statistics, with mean subtraction and division by standard deviation applied consistently during training and inference. The output layer uses a sigmoid activation function to produce values in the range $[0, 1]$, which are then rescaled to $[0, 20]$ using learned normalization parameters to match the target \lensfit weight range. The model is trained using mean absolute error (MAE) as the loss function rather than mean squared error, as MAE is less sensitive to outliers and better suited for the heavily skewed weight distribution. The \texttt{Adam} optimizer \citep{kingma14} is configured with an initial learning rate of 0.0004 and weight decay of $10^{-4}$ for $L_2$ regularization. Several regularization techniques prevent overfitting: a dropout rate of 0.3, batch normalization, and a dynamic batch size scheduling strategy with optimizer restart. The system monitors validation loss for plateaus using a patience of 5 epochs. When a plateau is detected, the batch size increases by a factor of 4 from an initial 512 up to a maximum of 8192, while the learning rate decreases by the same factor. This strategy allows the model to escape local minima and explore different regions of the loss landscape. Training data is restricted to KiDS-1000 galaxies only, with 70\% used for training and 30\% for validation.

\end{document}